\documentclass[graybox, envcountchap]{svmult}

\usepackage{mathptmx}        
\usepackage{amsmath}
\usepackage{amssymb}
\usepackage{color}
\usepackage{helvet}          
\usepackage{courier}         
\usepackage{dirtree}

\usepackage{makeidx}        
\usepackage{graphicx}        
\usepackage{subfig}

\usepackage{multicol}        
\usepackage[bottom]{footmisc}

\usepackage{hyperref}        
\hypersetup{colorlinks=true,urlcolor=blue}

\usepackage[misc]{ifsym}

\makeindex             

\usepackage{soul}

\begin{document}
\renewcommand{\hbar}{\mathchar'26\mkern-9mu h}


\title*{
Exotic compact objects: \\
a recent numerical-relativity perspective}
\author{Miguel Bezares and Nicolas Sanchis-Gual}
\institute{Miguel Bezares (\Letter) \at 
Nottingham Centre of Gravity and
School of Mathematical Sciences, University of Nottingham,
University Park, Nottingham NG7 2RD, United Kingdom. \email{Miguel.Bezaresfigueroa@nottingham.ac.uk}
\and Nicolas Sanchis-Gual \at Departamento de Astronom\'ia y Astrof\'isica, Universitat de Val\`encia, Dr. Moliner 50, 46100, Burjassot (Valencia), Spain. \email{nicolas.sanchis@uv.es}}
%
%
\maketitle

\abstract{Beyond black holes and neutron stars, new hypothetical compact objects have been proposed as potential astrophysical entities. In general, their properties have not yet been fully explored or understood, nor has it been proven whether or not they exist in nature. They are the so-called {\it exotic compact objects}, theoretical equilibrium configurations in the strong regime of gravity that involve new exotic physical phenomena deeply related to fundamental questions of theoretical physics (e.g., the nature of dark matter, the formation of singularities, or the presence of horizons). Among these exotic objects, there are those that require the existence of new fields and particles beyond the Standard Model, such as boson stars and ultralight bosons; those that seek to describe the dense equation of state of neutron stars as an even more extreme state of matter made up of free quarks; or those that exhibit additional properties related to extensions of General Relativity or even quantum gravity. However, in order to move from theoretical objects to astrophysical objects, their dynamics and stability must first be assessed by performing numerical-relativity simulations under the premise that solutions that are unstable on dynamical timescales will never completely form, being, at most, a transient state that will not be able to play any astrophysical role. Furthermore, numerical simulations also make it possible to extract the gravitational radiation from relevant astrophysical scenarios, such as the collapse of exotic stars or binary mergers, which could then be compared with current and upcoming LIGO-Virgo-KAGRA gravitational-wave detections.
}


\section{Introduction}
\label{sec:1}

Multimessenger observations involving gravitational wave detections by the LIGO-Virgo-KAGRA collaboration~\cite{GWTC1_PRX,abbott2021gwtc2,GWTC2.1,Populations_GWTC3} together with the imaging of the shadow of the supermassive black holes at the centre of Andromeda M87~\cite{akiyama2019first} and the Milky Way~\cite{akiyama2022first} galaxies by the Event Horizon Telescope are key to unveil the underlying theory of gravity and understand the true nature of the most compact objects in the Universe: black holes and neutron stars. Although the data so far are in excellent agreement with the Kerr hypothesis, which assumes that each astrophysical black hole in the Universe is described by the Kerr metric from General Relativity, it is worth asking if alternative, non-conventional models of compact objects~\cite{herdeiro2023black}, i.e., theoretical exotic compact objects that also involve strong gravitational fields and could mimic some key observable phenomenology of neutron stars and black holes~\cite{Olivares:2018abq,Herdeiro:2021lwl}, also populate the Cosmos. 

Exotic compact objects may be linked to relevant aspects in many areas of fundamental physics~\cite{Cardoso:2019rvt}: 
\begin{itemize}
    \item the nature of dark matter (bosonic stars, bosonic fields, Kerr black holes with bosonic hair);
    \item the equation of state of dense matter in the interior of neutron stars (anisotropic stars, strange stars, fermion-boson stars, scalarized neutron stars);
    \item new properties of black holes related to the presence (or not) of a horizon and singularities (fuzzballs, wormholes, gravastars, baby universes in $f(R)$ theories),
    \item modified theories of gravity (scalarized black holes, boson stars in $f(R)$ gravity).
\end{itemize}    
While many compact objects have been proposed in the literature, 
in order to consider them as true viable astrophysical candidates beyond standard astrophysical objects, such models would have to meet a list of requirements~\cite{herdeiro2023black}:
\begin{svgraybox}\begin{enumerate}
    \item to be described within a sound and well-motivated physical theory (for instance, boson stars in the Einstein-Klein-Gordon system);
    \item to be sufficiently stable against small perturbations and long-lived;
    \item to have a dynamical formation mechanism.
\end{enumerate}
\end{svgraybox}
\noindent These requisites, and in particular the last two, allow us to define the concept of dynamical robustness as a criterion for astrophysical viability. Therefore, any exotic compact object aiming to be astrophysically relevant should satisfy these benchmarks. Compact objects that become unstable too quickly or cannot be formed from non-fine-tuned initial conditions would not survive in our unrestful Universe. 

Research on highly dynamic astrophysical systems in the strong field regime of gravity relies on numerical relativity~\cite{alcubierre2008introduction,gourgoulhon20123+} to solve Einstein's field equations in a hyperbolic framework coupled to the matter field or fields under study. This means that numerical simulations can be performed to assess the stability and dynamical formation of exotic compact objects. For instance, rotating objects may form in binary mergers of non-spinning configurations. Moreover, once the viability of a particular exotic compact model has been established, such solutions can be studied in more relevant astrophysical scenarios in which gravitational-wave emission is present. Obtaining gravitational waveforms from exotic compact objects and comparing them with real gravitational wave events could lead to their discovery or to discarding them~\cite{Bustillo:2020syj}.

This chapter will be devoted to discussing the (in)stability and formation channel of different theoretical compact objects determined using numerical-relativity simulations and their associated gravitational wave signature, if any. In Section~\ref{Section1}, we will present recent results on the stability of isolated single exotic objects in General Relativity and comment on the implications of the findings. In Section~\ref{Section2}, we will discuss some examples of numerical simulations of compact objects in modified theories of gravity. Finally, in Section~\ref{Section3}, we will examine the binary merger scenarios and the gravitational radiation emitted by these objects. Throughout this chapter, Roman letters from the beginning of the alphabet $a, b, c, \ldots$ denote space-time indices ranging from $0$ to $3$, while letters near the middle $i, j, k, \ldots$ range from 1 to 3, denoting spatial indices.

\section{Exotic matter in General Relativity}\label{Section1}
Dark matter and its properties are one of the fundamental open questions in astrophysics, cosmology, and particle physics. What massive particle(s) that mainly interacts gravitationally, if there is indeed any,  is responsible for the multiple dark-matter gravitational effects we see in galaxies~\cite{rubin1978extended}? Although there are many candidates, such as WIMPS (Weakly Interacting Massive Particle), axions and axion-like particles, new fundamental bosons, and sterile neutrinos, there is no evidence of the existence of any of them. Here, we will focus on ultralight bosons with masses in the range $10^{-22}<m<10^{-10}$ eV, as an example of a potential dark matter particle~\cite{Arvanitaki:2009fg}. These particles could condense into a new type of compact object known as a bosonic star, which can be described as a self-gravitating Bose-Einstein condensate, and also grow through superradiance to form clouds around rotating black holes~\cite{Brito:2015oca}. 

Besides dark matter, another open question in astrophysics is related to the interior of extremely dense neutron stars: What is the equation-of-state of dense matter? Does matter suffer a phase transition under such extreme conditions and is better described as made of quarks rather than neutrons? Furthermore, these exotic objects allow us, for example, to circumvent the Buchdahl limit and the existence of anisotropies in neutron stars. They also make possible the study of open problems in theoretical physics, such as the information-loss paradox and quantum description of the horizon.

\begin{svgraybox}In this section we will briefly present some compelling models of isolated compact objects described by exotic matter and discuss the numerical simulation that attempt to determine their stability and dynamical formation.\end{svgraybox}

\subsection{Boson and Proca stars}
A boson star is a collection of bosons in their ground state that composes a self-gravitating macroscopic Bose-Einstein condensate described by a complex scalar field. They are horizonless, stationary, everywhere regular, solitonic solutions of the Einstein-(complex, massive) Klein-Gordon system~\cite{Kaup:1968zz,Ruffini:1969qy,Jetzer:1991jr}. Its vector cousins are known as Proca stars and are solutions of the Einstein-(complex) Proca system~\cite{Brito:2015pxa,Herdeiro:2023wqf}.
They are described by the Einstein-Klein-Gordon and Einstein-Proca actions (with units $c=1=\hbar$ and $4\pi G=1$):
\begin{eqnarray}
\label{action}
\mathcal{S}=\int d^4 x \sqrt{-g} 
\left [
\frac{R}{4} 
+
\mathcal{L}_{(s)}
\right] \ ,
\end{eqnarray}
with the two matter Lagrangians being:
\begin{eqnarray}
\label{lagrangians}
&&\mathcal{L}_{(0)}= - g^{ab}\partial_a\Phi^{*}\partial_b \Phi - \mu^2 \Phi^{*}\Phi - V_{\rm{int}}(\Phi,\Phi^{*}) \ , \\ 
&&\mathcal{L}_{(1)}= -\frac{1}{4}\mathcal{F}^{*}_{ab}{\mathcal{F}}^{ab}
-\frac{1}{2}\mu^2\mathcal{A}^{*}_a {\mathcal{A}}^a \ .
\end{eqnarray} 
$\Phi$ is a complex scalar field, $V_{\rm{int}}$ is a high-order self-interaction potential, and $\mathcal{A}$ is a complex 4-potential, with field strength $\mathcal{F} =d\mathcal{A}$. The asterisk $*$ indicates complex conjugate and $\mu$ is the field's mass. The complex bosonic field has a harmonic dependence and oscillates in time with a well-defined frequency $\omega$, while at the level of the stress-energy tensor, the configuration is time-independent, leading to a stationary spacetime. The scalar field ansatz for boson stars is given by:
\begin{equation}\label{eq:ansatz}
\Phi(t,r,\theta,\phi) = \Phi_0(r,\theta)e^{i(m\phi-wt)}.
\end{equation}
where $m$ is the azimuthal winding number, which is an integer that can take both negative and positive values $m=0,\pm1,\pm2$. As for fermion stars, such as white dwarfs and neutron stars, bosonic stars have a maximum mass as well, which separates the stable and unstable branches, but it can vary from one to millions of solar masses, depending on the bosonic particle mass $\mu$~\cite{Berti:2006qt}:
\begin{equation}
M_{\rm{max}}=\mathcal{M}_{\rm{model}}\frac{M_{\rm{Planck}}^2}{\mu},
\end{equation}
where $\mathcal{M}_{\rm{model}}$ is a parameter that depends on the specific bosonic star model (with $\mathcal{M}_{\rm{model}}=0.633$ for non-spinning scalar mini-boson stars\footnote{Boson stars with only the mass term in~Eq.~\eqref{lagrangians}~[$V_{\rm{int}}=0$] are usually referred to as \it{mini}-boson stars.}) and $M_{\rm{Planck}}$ is the Planck mass. This feature is what makes them especially captivating, both from an astrophysical and a particle physics point of view. A lighter boson leads to more massive stars, beautifully connecting different mass and length scales. Hence, there is a need for ultralight bosons, with masses below $10^{-10}$ eV to reach astrophysical mass ranges. In this regard, fundamental bosonic fields minimally coupled to gravity are widely used in cosmological models to explain the nature of dark matter~\cite{Matos:2000ng}, as a consequence of String Theory~\cite{Arvanitaki:2009fg} or to extend the Standard Model of particle physics~\cite{Freitas:2021cfi}, with axions or dark photons.

Therefore, bosonic stars are exotic compact objects well motivated by particle physics and General Relativity (point 1). Nonetheless, these solutions could also be used as toy models of more involved exotic compact objects or different types of dark matter. The existence and discovery of the Higgs boson gave a fresh impetus to this research. However, as stated above, to truly contemplate them as potential astrophysical objects, numerical studies need to be undertaken to fully determine their stability and find the mechanisms of their dynamical formation. Within spherical symmetry, scalar boson star models with and without self-interaction terms in their potential have been extensively studied over the last few decades both analytically~\cite{Gleiser:1988rq,lee1989stability,Schunck:2003kk,Santos:2024vdm} and numerically in dynamical time evolutions~\cite{Liebling:2012fv}. They have been shown to be indeed stable against (sufficiently small) linear perturbations and to form in highly dynamical scenarios through the so-called gravitational cooling mechanism~\cite{Seidel:1993zk,DiGiovanni:2018bvo} (points 2 and 3). As an illustrative example, the evolution equations for a free scalar field (without self-interactions) are shown:
\begin{eqnarray}
\partial_{t} \Phi & =& - \alpha \Pi + \mathcal{L}_{\beta} \Phi\
                \,, \label{eq:dtPhi} \\
  \partial_{t} \Pi &  =& \alpha \left( K \Pi - \gamma^{ij} D_i \partial_j \Phi
                  +  \mu^2 \Phi \right)  -  \gamma^{ij} \partial_i \alpha \partial_j \Phi
                       + \mathcal{L}_{\beta} \Pi \,, \label{eq:dtKphi}
\end{eqnarray}
where $\alpha$, $\beta$, $\gamma^{ij}$, and $K$ are the lapse function, the shift vector, the three-dimensional spatial metric, and the trace of the extrinsic curvature $K_{ij}$, respectively, which are variables that come from the 3+1 decomposition of Einstein equations; $\mathcal{L}_\beta$ is the Lie derivative with respect to the shift vector; and $\Pi$ is the conjugated momenta to the scalar field, defined as $\Pi=-(\partial_t - \mathcal{L}_\beta)\Phi/\alpha.$ These equations need to be solved coupled to a 3+1 evolution formalism for the gravitational field, such as the Baumgarte-Shapiro-Shibata-Nakamura formulation~\cite{Shibata:1995we,Baumgarte:1998te} or the conformal and covariant Z4 formulation (CCZ4)~\cite{alic}. Spherically symmetric non-spinning scalar boson stars are dynamically robust and well-motivated candidates to be (part of) dark matter.

\begin{figure}[h!]
\centering
\includegraphics[width=0.24\linewidth]{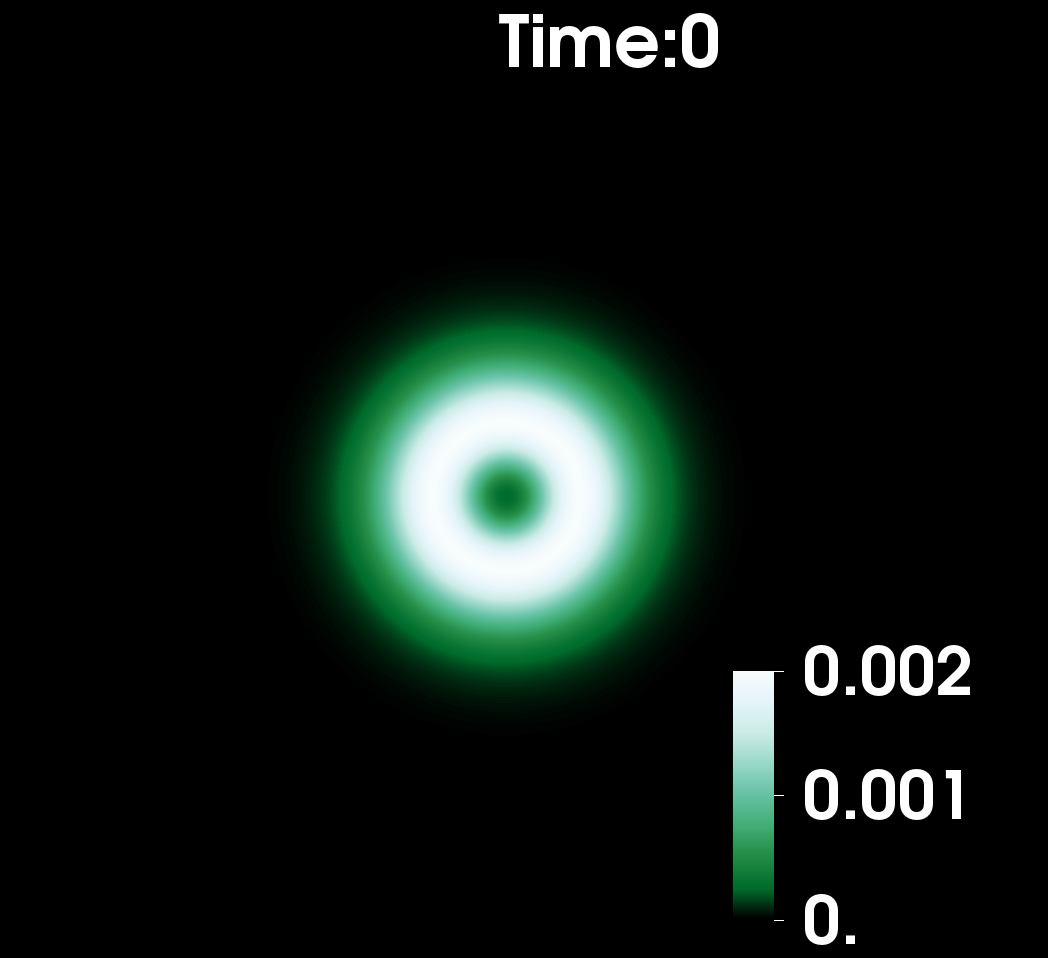}\hspace{-0.01\linewidth}
\includegraphics[width=0.24\linewidth]{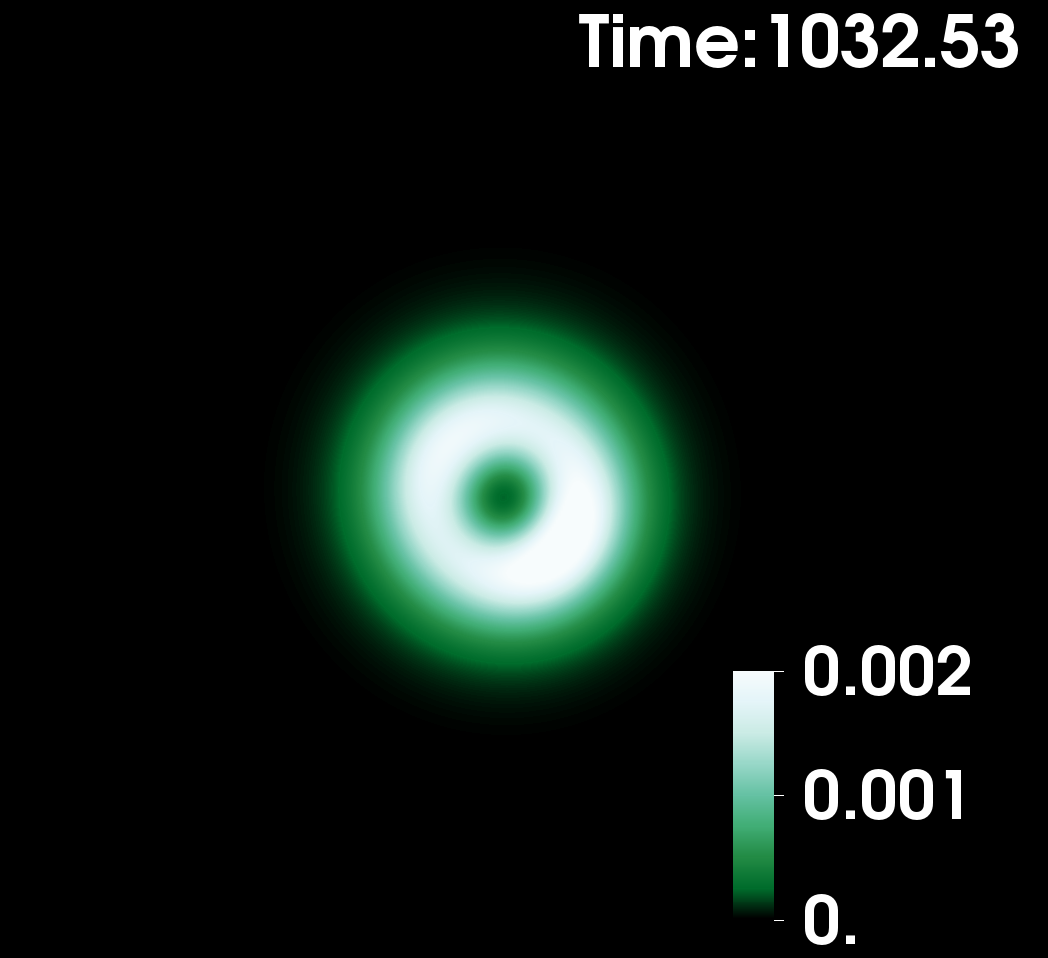}\hspace{-0.01\linewidth}
\includegraphics[width=0.24\linewidth]{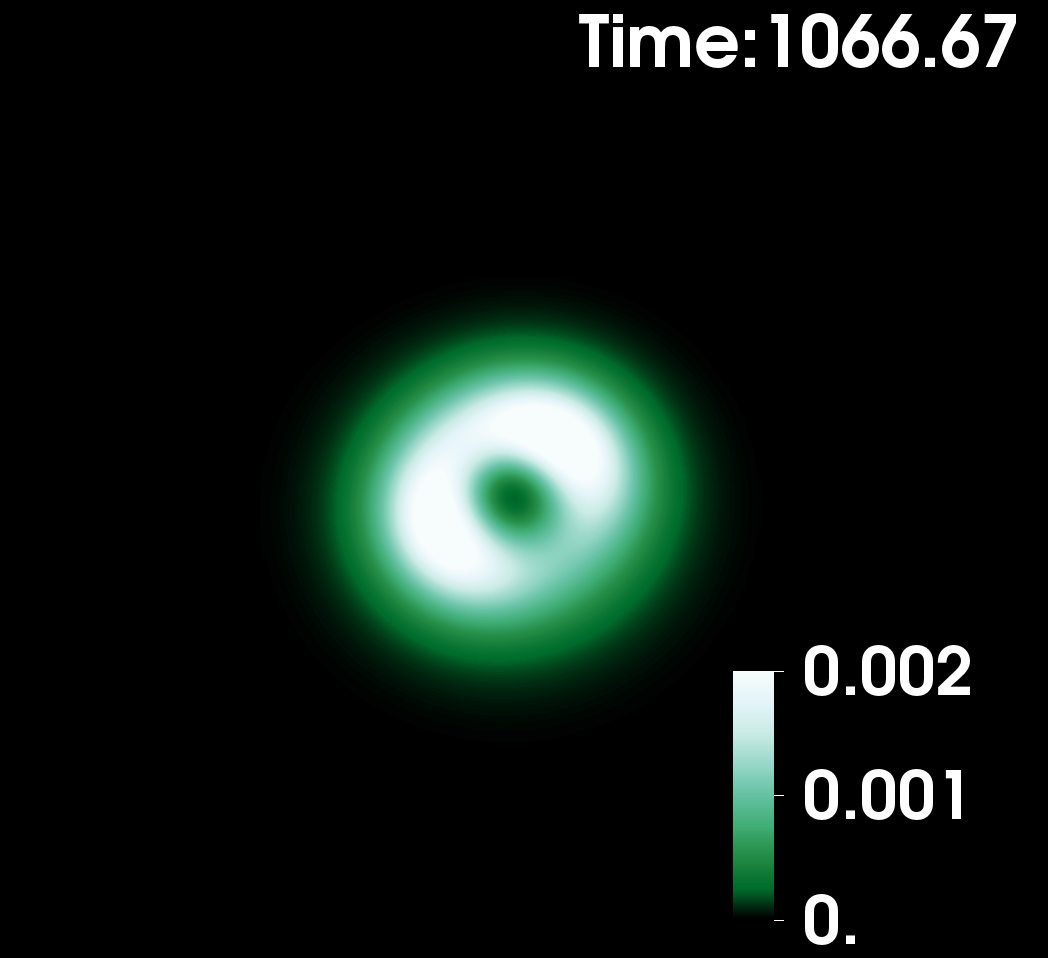}\hspace{-0.01\linewidth}
\includegraphics[width=0.24\linewidth]{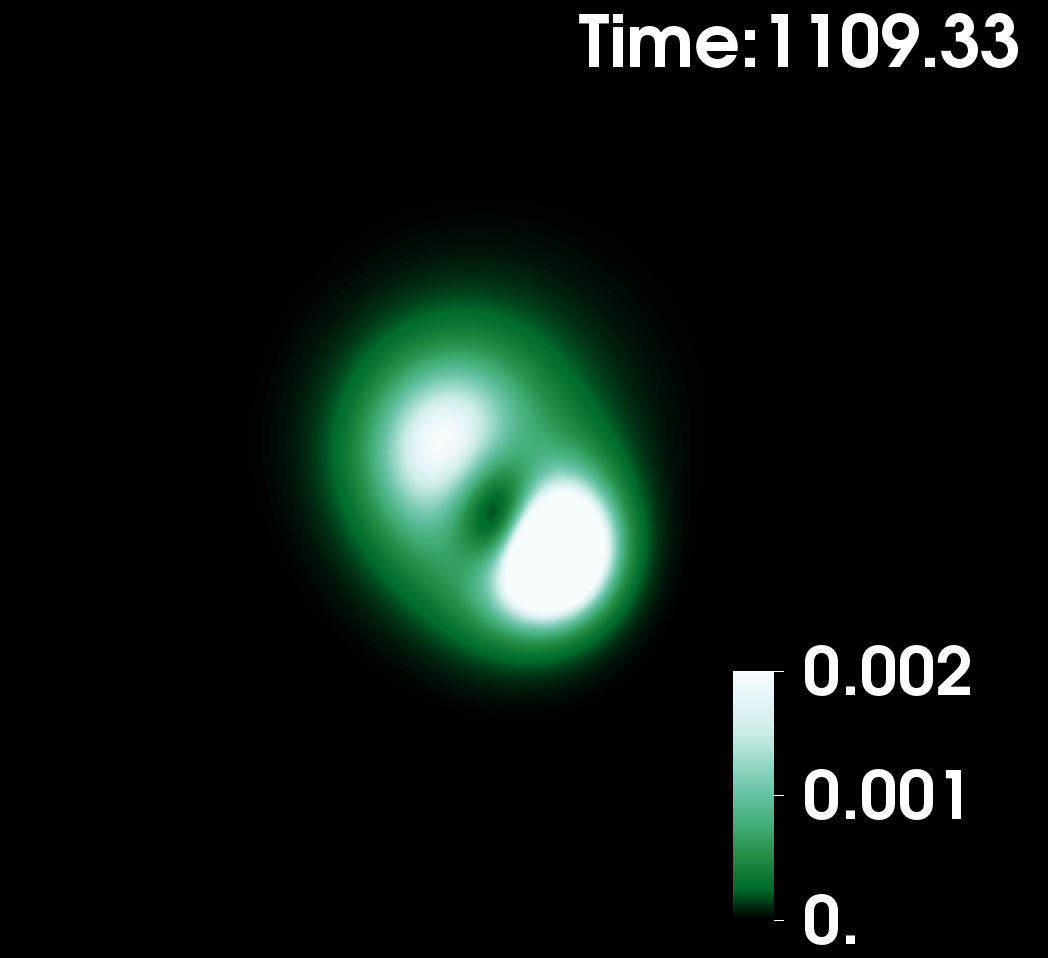}\hspace{0.02\linewidth}\\
\includegraphics[width=0.24\linewidth]{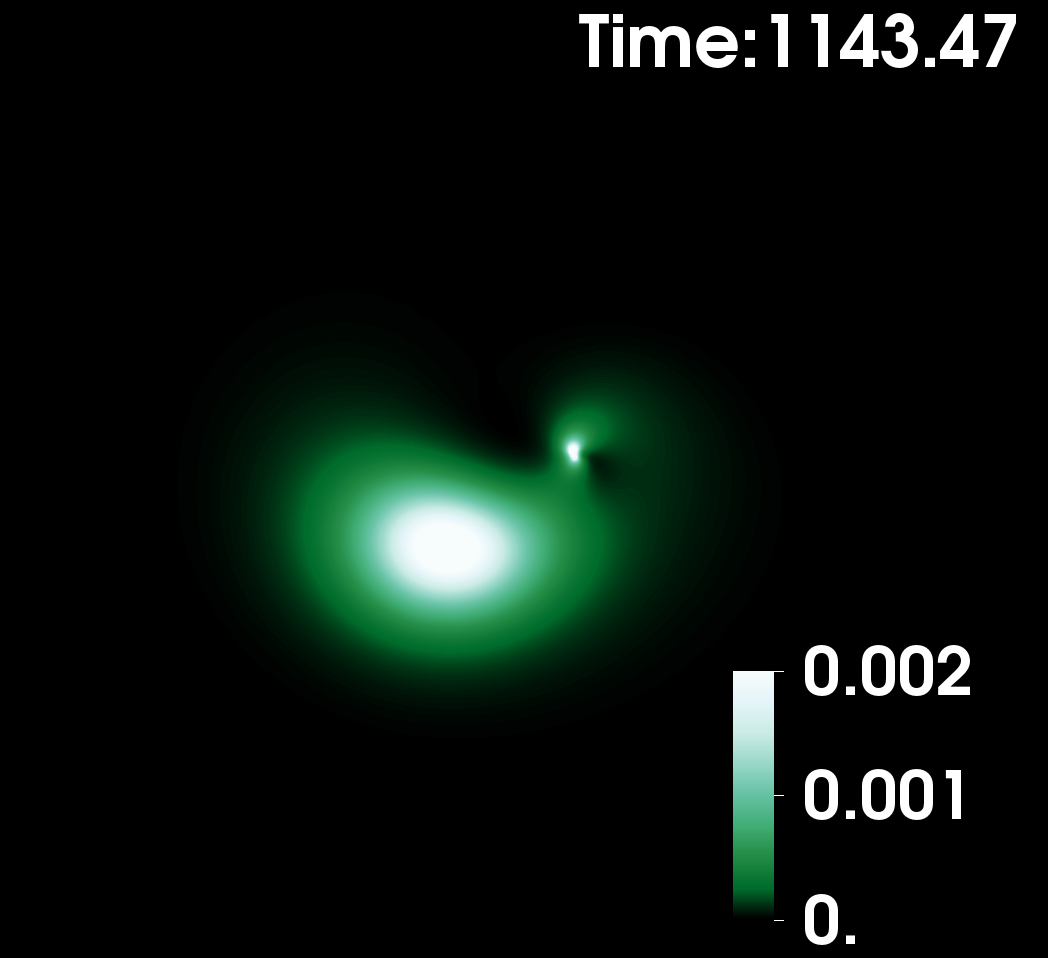}\hspace{-0.01\linewidth}
\includegraphics[width=0.24\linewidth]{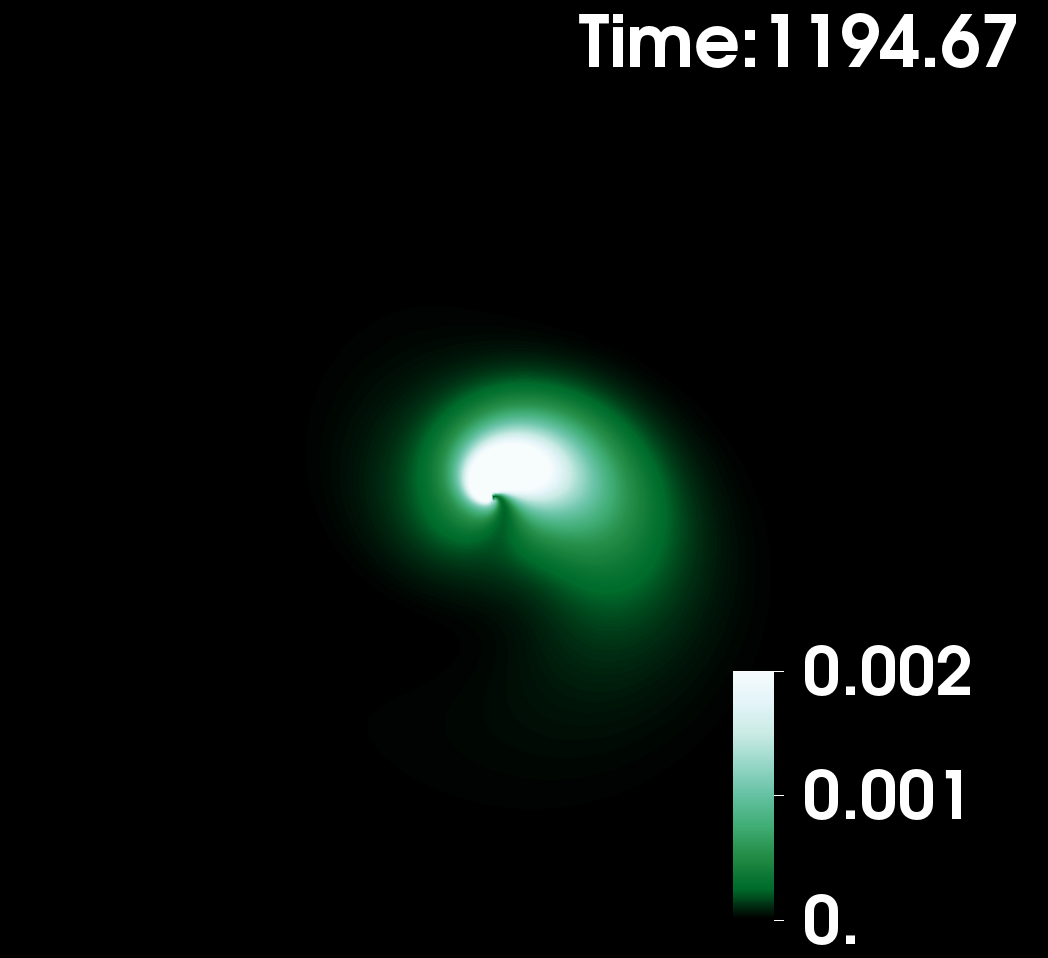}\hspace{-0.01\linewidth}
\includegraphics[width=0.24\linewidth]{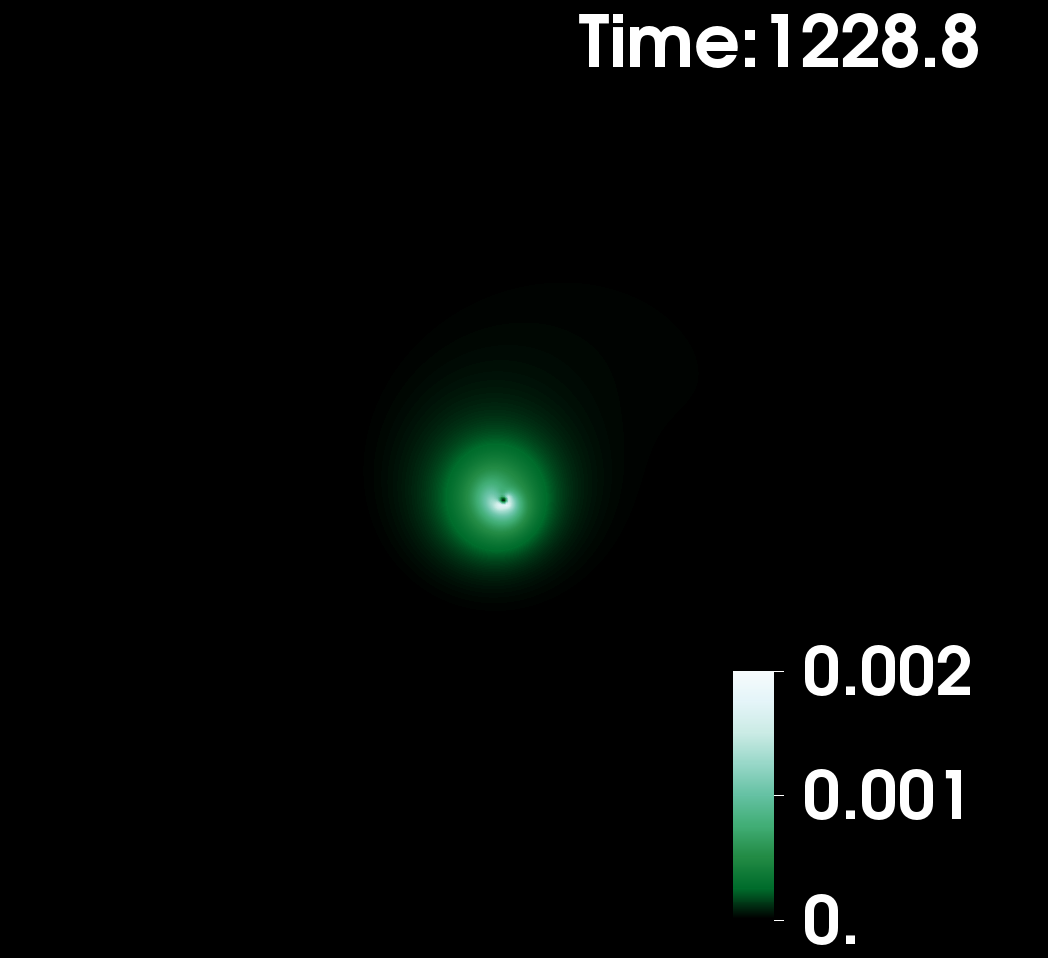}\hspace{-0.01\linewidth}
\includegraphics[width=0.24\linewidth]{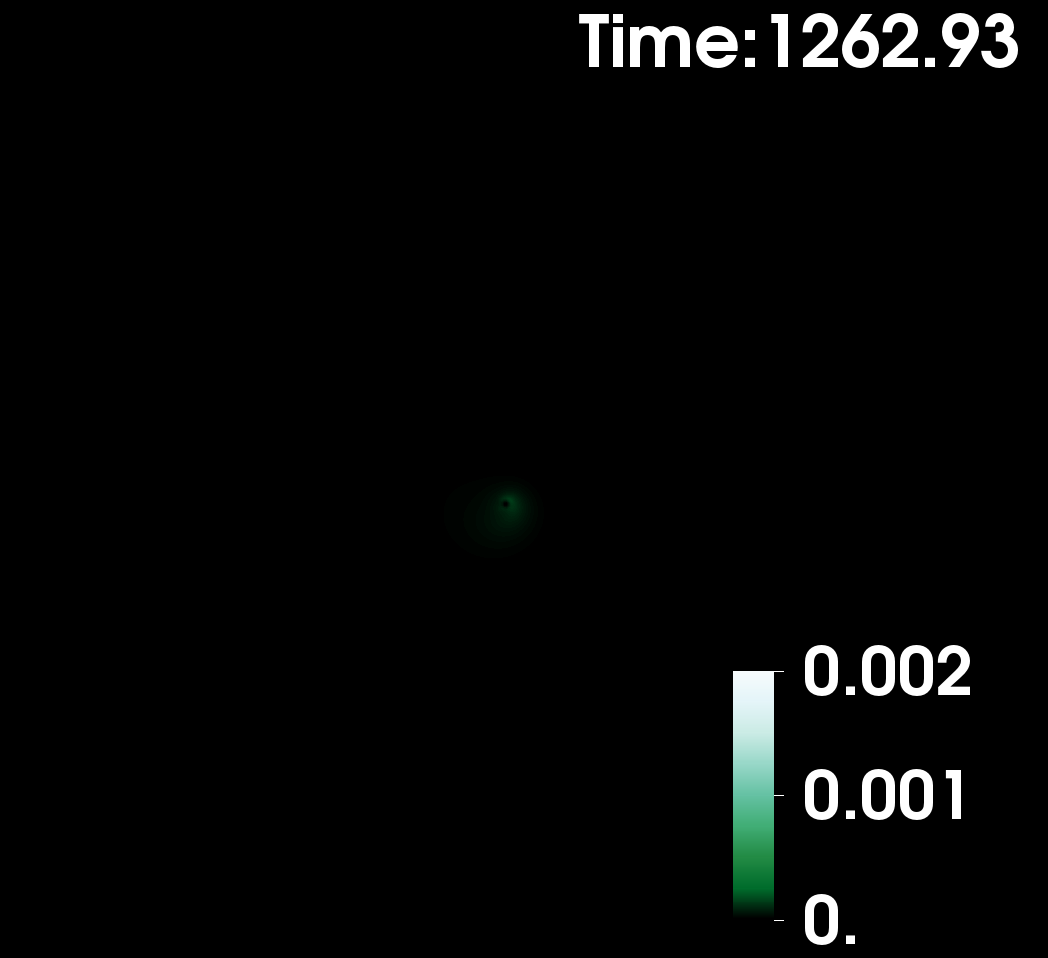}\hspace{0.02\linewidth}\\
\caption{ {\em Spinning bosonic stars.} Time evolution of the energy density of a scalar spinning boson star. Time runs left to right, $1^{st}$ to $2^{nd}$ row. Extracted from~\cite{Sanchis-Gual:2019ljs}.}
\label{fig:121_boson}
\end{figure}

\begin{figure}
\centering
\includegraphics[width=0.23\linewidth]{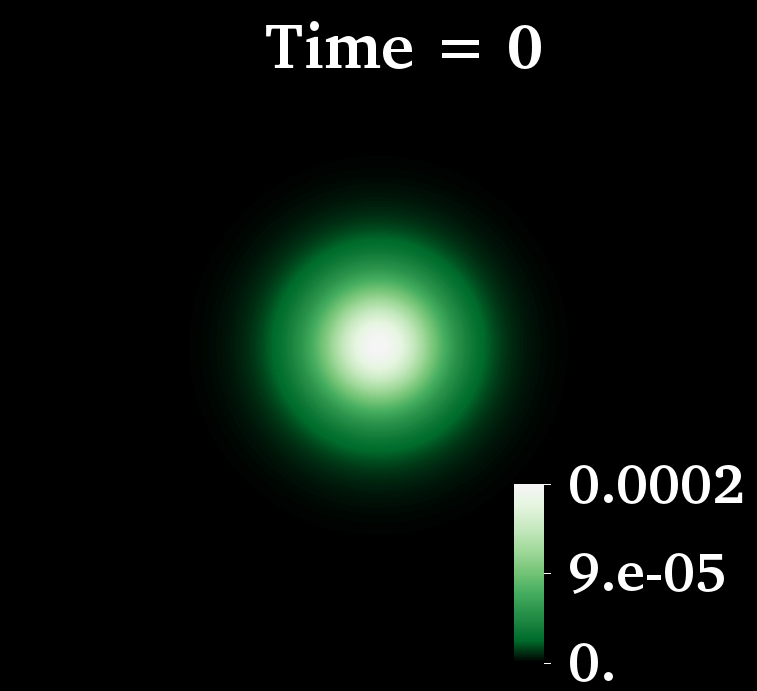}\hspace{-0.01\linewidth}
\includegraphics[width=0.23\linewidth]{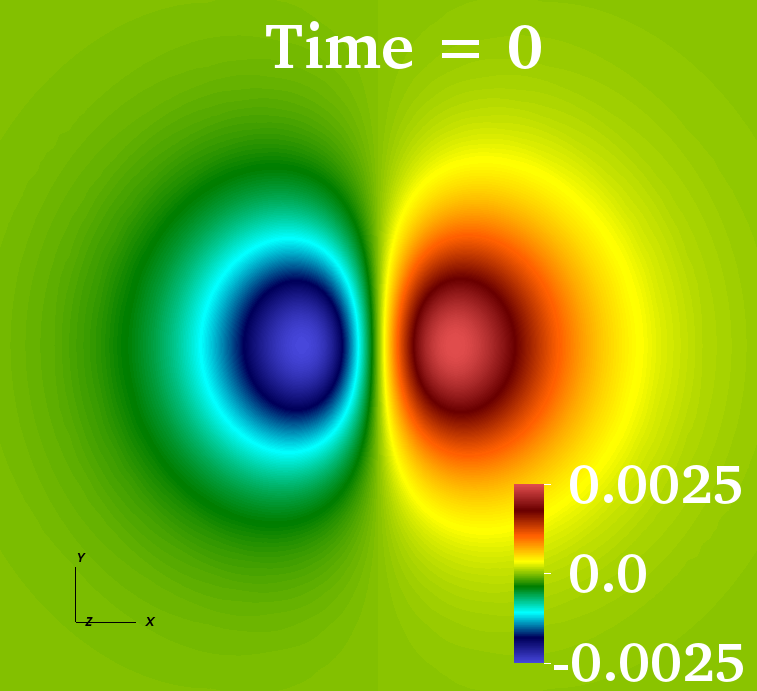}\hspace{-0.01\linewidth}
\includegraphics[width=0.23\linewidth]{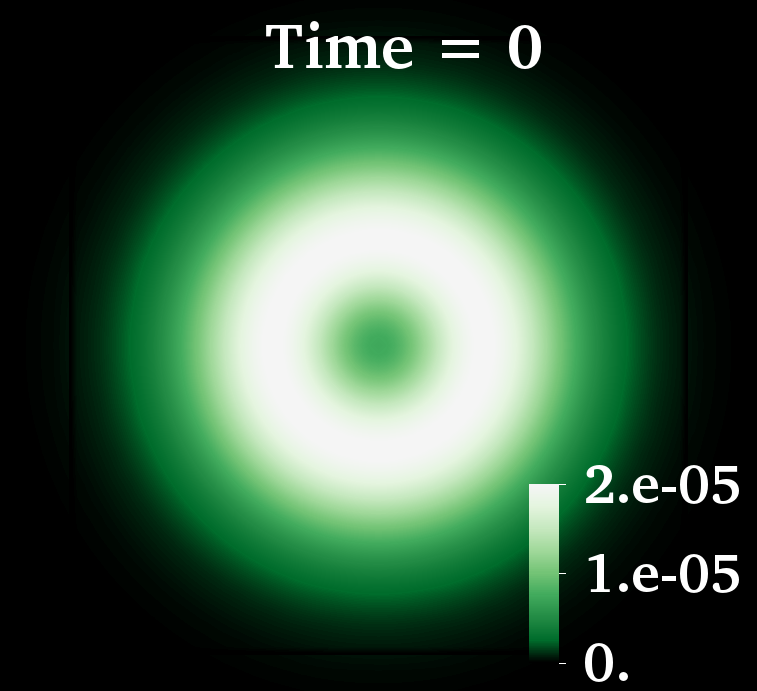}\hspace{-0.01\linewidth}
\includegraphics[width=0.23\linewidth]{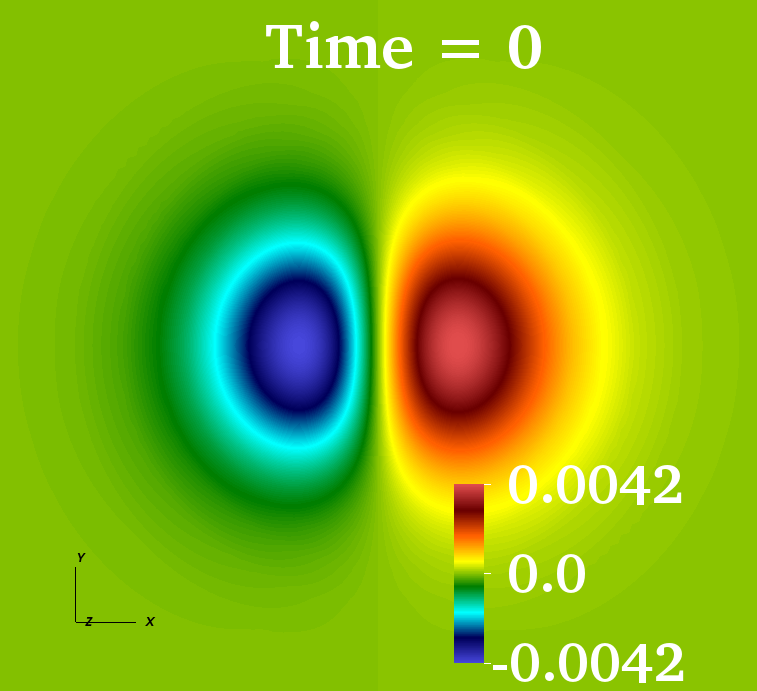}\hspace{-0.01\linewidth}\\
\includegraphics[width=0.23\linewidth]{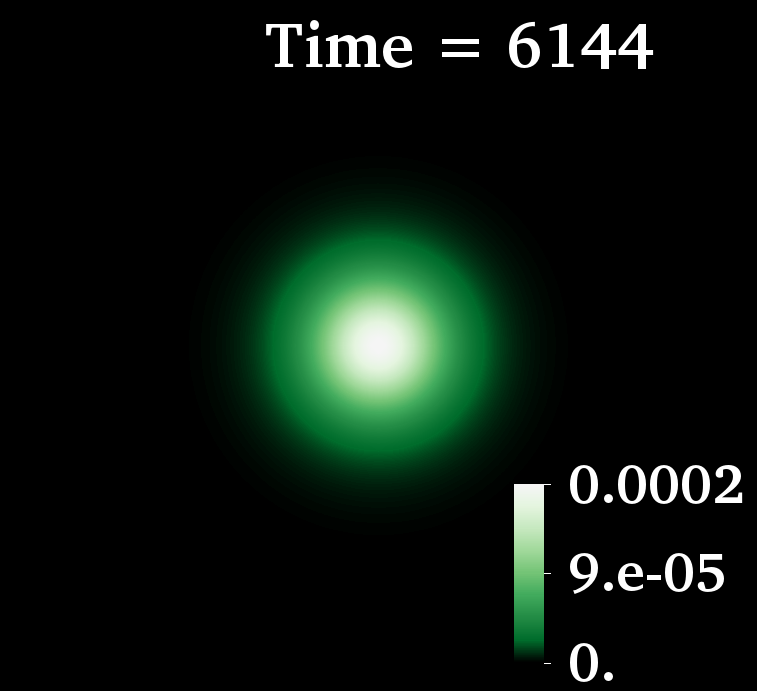}\hspace{-0.01\linewidth}
\includegraphics[width=0.23\linewidth]{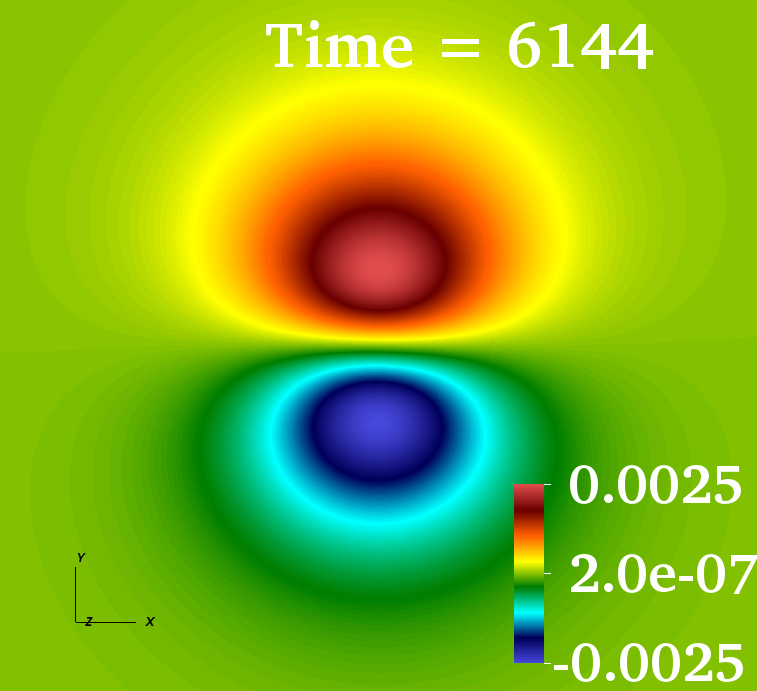}\hspace{-0.01\linewidth}
\includegraphics[width=0.23\linewidth]{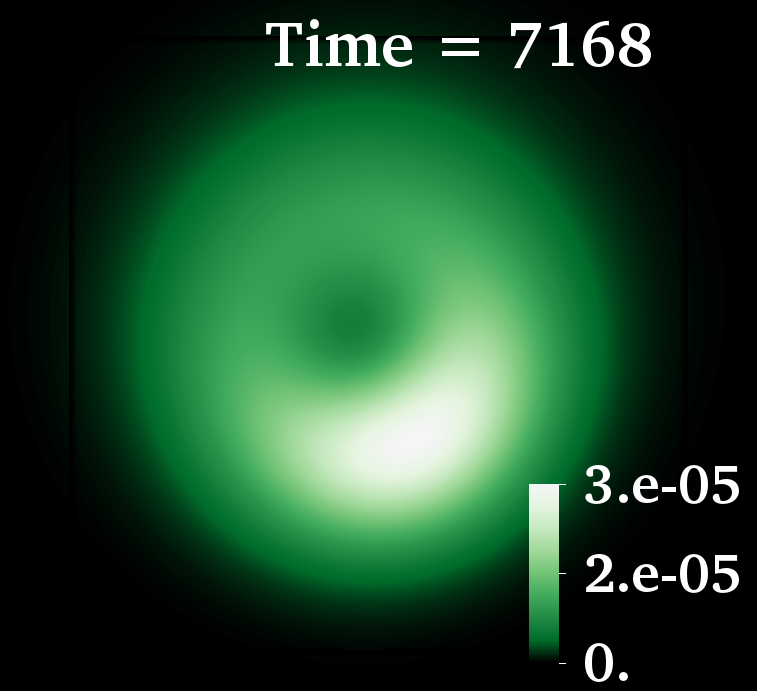}\hspace{-0.01\linewidth}
\includegraphics[width=0.23\linewidth]{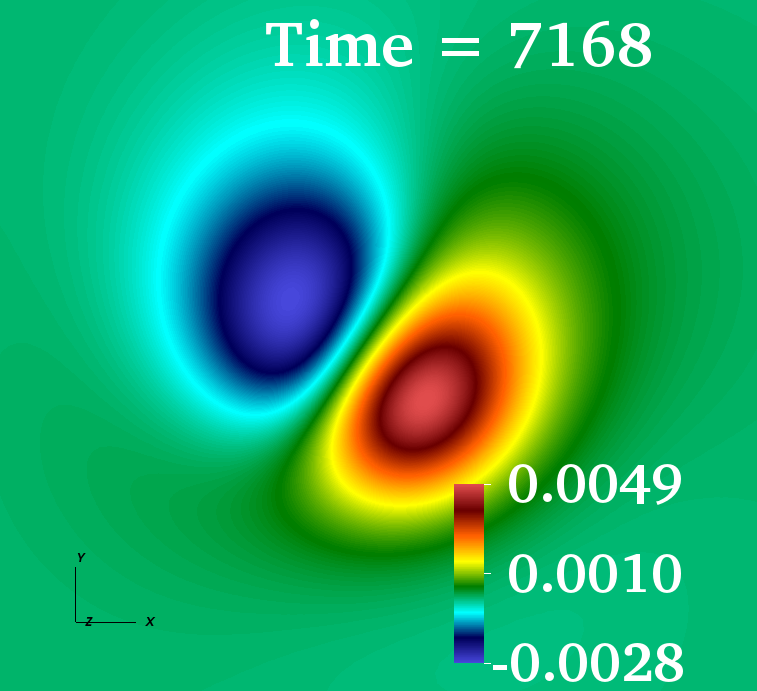}\hspace{-0.01\linewidth}\\
\includegraphics[width=0.23\linewidth]{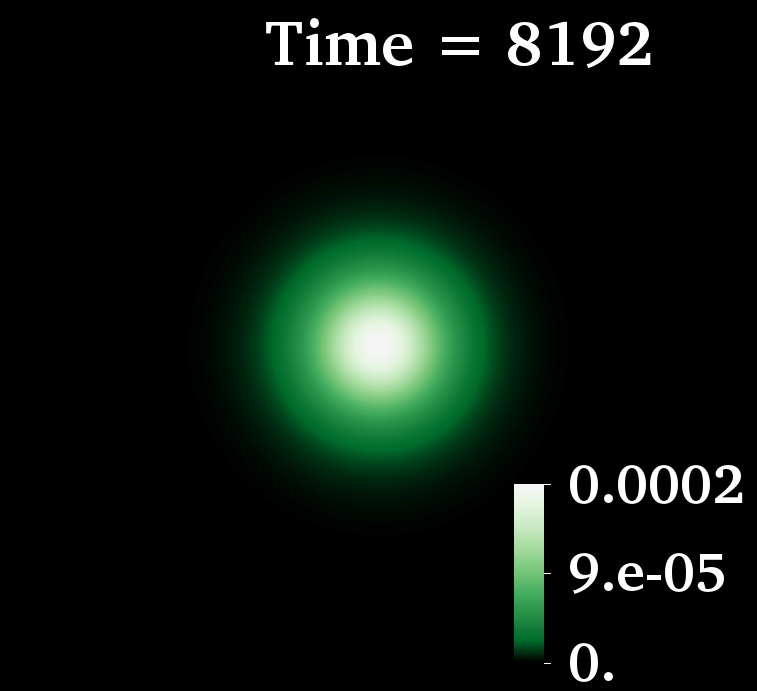}\hspace{-0.01\linewidth}
\includegraphics[width=0.23\linewidth]{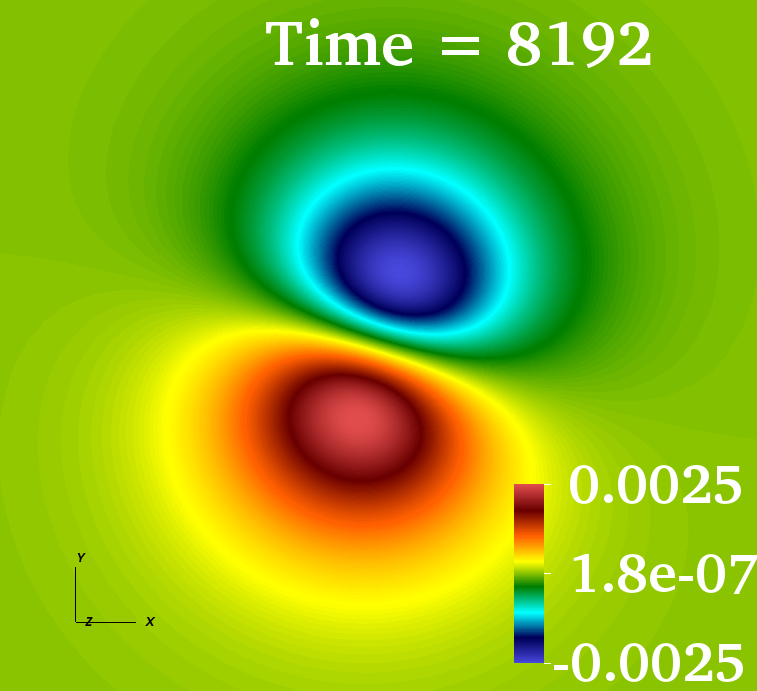}\hspace{-0.01\linewidth}
\includegraphics[width=0.23\linewidth]{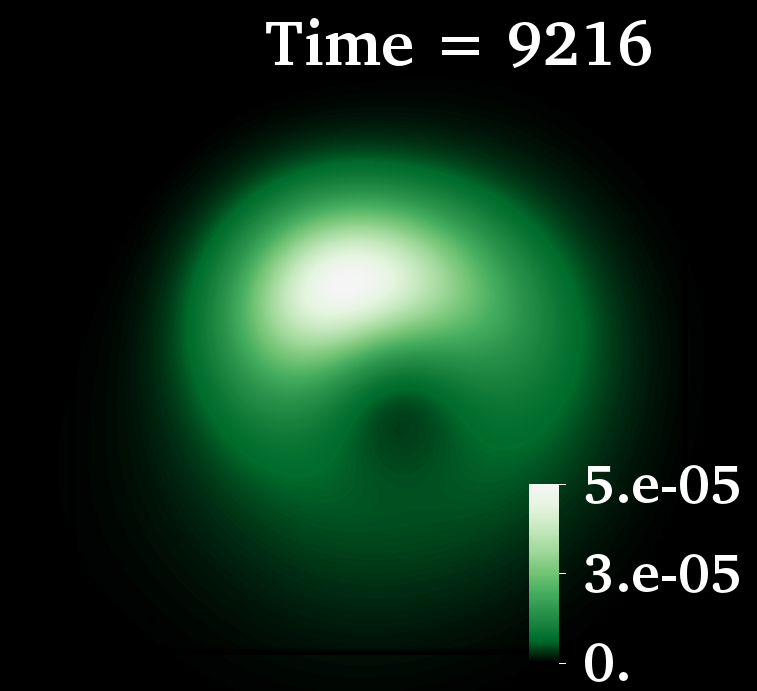}\hspace{-0.01\linewidth}
\includegraphics[width=0.23\linewidth]{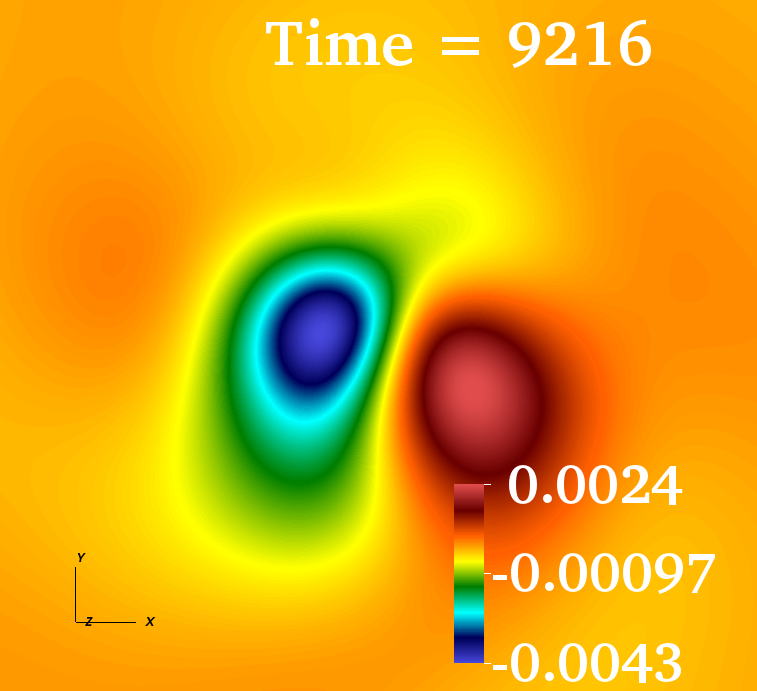}\hspace{-0.01\linewidth}\\
\includegraphics[width=0.23\linewidth]{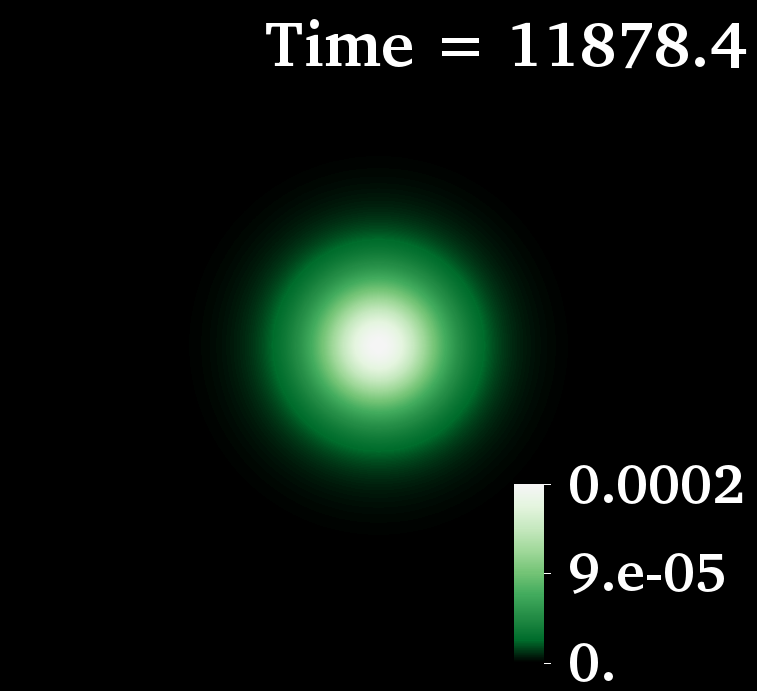}\hspace{-0.01\linewidth}
\includegraphics[width=0.23\linewidth]{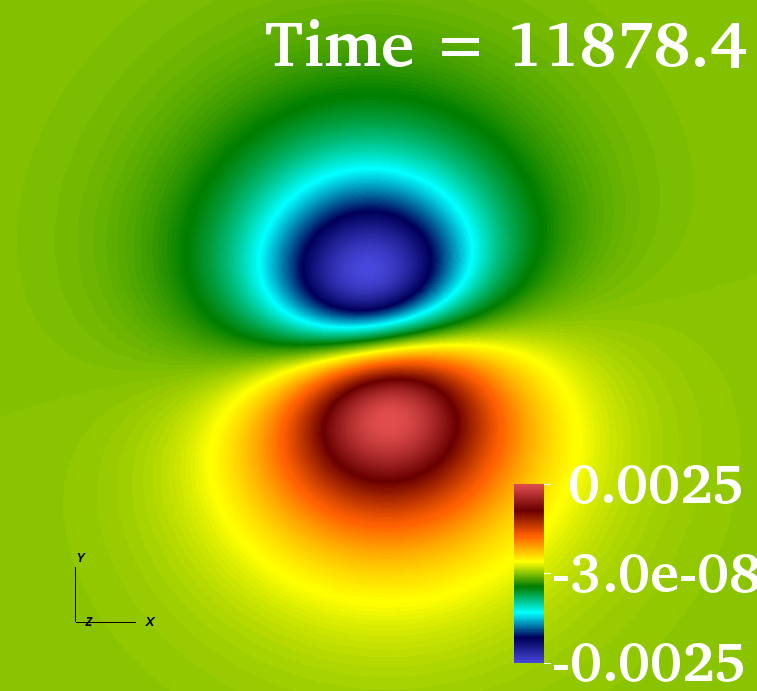}\hspace{-0.01\linewidth}
\includegraphics[width=0.23\linewidth]{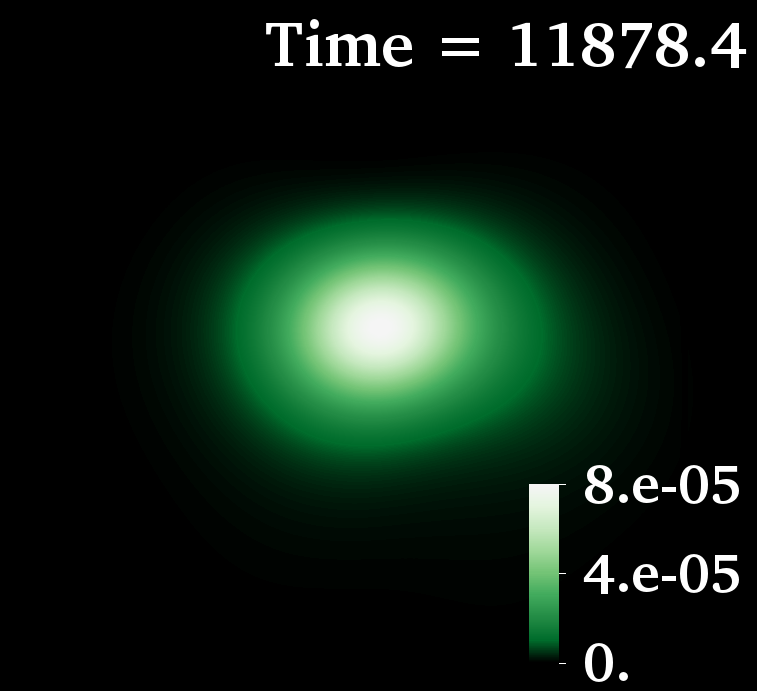}\hspace{-0.01\linewidth}
\includegraphics[width=0.23\linewidth]{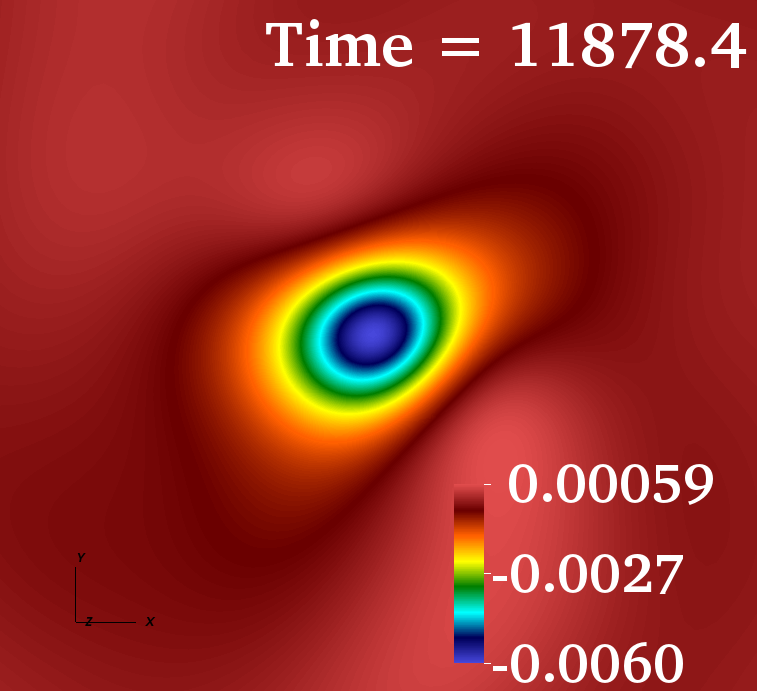}\hspace{-0.01\linewidth}\\
\caption{ {\em $\ell$-bosonic stars.} Time evolution of the energy density and the real part of the $\ell=m=1$ mode of scalar multi-field boson stars, corresponding a superposition of a static non-spinning $\ell=m=0$ and a rotating $\ell=m=1$ components, with $\omega_1=0.967$ (leftmost and left middle column), and with $\omega_1=0.974$ (right middle and rightmost column). Time runs from top to bottom. Extracted from~\cite{Sanchis-Gual:2021edp}.}
\label{fig:122_multi}
\end{figure}

Natural extensions of the spherical case have been proposed and studied in recent years.
\begin{itemize}
\item Rotating boson stars with $m\neq0$~\cite{Yoshida:1997qf};  numerical studies showed that rotating scalar mini-boson stars are unstable against non-axisymmetric perturbations~\cite{Sanchis-Gual:2019ljs,DiGiovanni:2020ror}, developing a bar-mode instability similar to that of differentially rotating neutron stars (see Figure~\ref{fig:121_boson}). These were unexpected results since rotating boson stars were believed to be stable according to catastrophe theory~\cite{Kleihaus:2011sx}. The decay to the spherical or lower angular momentum states is accompanied by a long-lived gravitational-wave emission~\cite{DiGiovanni:2020ror}.
\item Dipolar boson stars described by a $\ell=1$, $m=0$ mode and excited states with one or more nodes in the radial profile of the field~\cite{Jetzer:1989vs}; these solutions are unstable and decay dynamically to the fundamental nodeless spherically symmetric $ell=m=0$ configurations~\cite{Balakrishna:1997ej,Sanchis-Gual:2021edp,Ildefonso:2023qty}. 
\item Self-interactions with high-order terms in the potential; their inclusion leads to an increase in the maximum mass of bosonic stars~\cite{Colpi:1986ye} and they can mitigate some of the dynamical instabilities of these stars (e.g. rotating models~\cite{Siemonsen:2020hcg,Dmitriev:2021utv}, excited states~\cite{Sanchis-Gual:2021phr,Brito:2023fwr}, and dipolar boson stars~\cite{Ildefonso:2023qty} can be stabilized).
\item Multistate boson stars~\cite{Bernal:2009zy}, $\ell$-boson stars~\cite{Alcubierre:2018ahf,Jaramillo:2020rsv} and multifield boson stars~\cite{Guzman:2019gqc,Guzman:2021oma,Sanchis-Gual:2021edp}; these solutions are superpositions of boson stars in different states described by several complex scalar fields coupled to gravity that can modify the stability properties of isolated stars as shown in Fig.~\ref{fig:122_multi}, where the numerical evolution of multi-state stars demonstrated the quenching of instabilities.
\item Proca stars were proposed in 2016~\cite{Brito:2015pxa} as the vector counterparts of scalar boson stars. The configurations resemble very much the scalar case and share similar features. However, they have different dynamical stability properties. Spinning stars described by a $\ell=m=1$ mode have a spheroidal morphology and contrary to the rotating scalar case, they are dynamically robust (stability and dynamical formation)~\cite{Sanchis-Gual:2019ljs}. On the other hand, $\ell=m=0$ spherically symmetric stars do not constitute the fundamental ground state of solutions, according to recent numerical simulations~\cite{Herdeiro:2023wqf} they decay to other static configurations with axial symmetry but not spherical symmetry: these new solutions haven been dubbed as $\ell=1$, $m=0$ prolate Proca stars.
\item Bosonic stars with light rings; it has been argued that horizonless ultracompact objects with light rings could mimic part of the gravitational-wave ringdown signal in binary mergers~\cite{Cardoso:2016rao}. For this reason, observing gravitational wave ringdowns would not provide conclusive proof of the formation of horizons. However, such solutions must have at least two light rings, one stable and one unstable~\cite{Cunha:2017qtt}. The presence of the stable light ring could act as a trapping mechanism for null perturbations that could then trigger an instability~\cite{Cardoso:2014sna}. Ultracompact rotating bosonic stars (both scalar and Proca) indeed suffer a dynamical instability when light rings are present, leading to a decay to a less compact model or a collapse to a black hole, as shown in~\cite{Cunha:2022gde}, suggesting that ultracompact objects with light rings might be unstable.
\end{itemize}

These examples illustrate why it is so important to investigate the stability of exotic compact objects, especially through numerical simulations. Rotating scalar boson stars and static, spherically symmetric Proca stars on the stable branch were definitely \textit{expected} to be stable. And in a sense \textit{they are stable}, since they are on the stable branch against linear radial perturbations and do not suffer the same fate as the configurations on the unstable branch. Still, these bosonic stars are prone to other types of instabilities, in particular, non-axisymmetric instabilities that lead to their decay. In Table~\ref{tab:1}, we give a summary of the stability of different classes of boson and Proca stars for configurations in the branch before the maximum mass model.

\begin{table}
\caption{ {\em Bosonic stars.} Non-radial (in)stability of mini-bosonic stars on the stable branch.}
\label{tab:1}       
\begin{tabular}
{p{2cm}p{2.4cm}p{2cm}p{2cm}p{2cm}}
\hline\noalign{\smallskip}
 & Static & Static & Rotating & Rotating  \\
Classes & $\ell=m=0$ & $\ell=1$, $m=0$ & $\ell=1$, $m=\pm1$ & $\ell=2$, $m=\pm2$  \\
\noalign{\smallskip}\svhline\noalign{\smallskip}
Scalar & Stable & Unstable & Unstable & Unstable\\
Proca & Unstable & Stable & Stable & Unstable\\
\noalign{\smallskip}\hline\noalign{\smallskip}
\end{tabular}
\end{table}

\subsection{Fermion-boson stars}\label{fbs}

Fermion-boson stars~\cite{Henriques:1989ar,Henriques:1990xg} are compact stars modelled as a superposition of a fermion and a boson star, respectively. The fermionic part can be described as a perfect fluid. Hence, the energy-momentum tensor for a fermion-boson star is 
\begin{eqnarray}
T_{ab} &=& T^{\Phi}_{ab} + T^{M}_{ab}~,
\label{FB_tensor}
\end{eqnarray}
where $T^{\Phi}_{ab}$ is the energy-momentum tensor associated to a (complex) scalar field and $T^{M}_{ab}$ corresponds to a perfect fluid. Defined as follows  
\begin{eqnarray}
T^{\Phi}_{ab} &=& \nabla_{a} \Phi \nabla_{b} \Phi^{*} + \nabla_{a} \Phi^{*}\nabla_{b} \Phi \nonumber\\ 
&&\qquad - \quad g_{ab} \left[ \nabla^{c} \Phi \nabla_{c} \Phi^{*}  + \mu^2 \Phi^{*}\Phi + V_{\rm{int}}(\Phi,\Phi^{*}) \right],\\
T^{M}_{ab} &=& [\rho(1+\epsilon)+P] u_{a}u_{b} + P\,g_{ab}~,   
\end{eqnarray}
where $\Phi^{*}$ is the complex conjugate of $\Phi,$  $\rho$ is the rest mass density, $\epsilon$ is the internal energy, $P$ is the pressure, which is given by an equation of state, that relates the pressure with the other fluid quantities, typically the rest mass density and the internal energy, and $u^{a}$ is the fluid four-velocity. The evolution equations of fermion-boson stars are then governed by the Einstein-Klein-Gordon-Euler system of equations, which are 
\begin{eqnarray}
G_{ab} &=&  8\pi\,(T^{\Phi}_{ab} + T^{M}_{ab}) ~,\label{FB_EE}\\
g^{ab} \nabla_{a} \nabla_{b} \Phi &=& \left( \mu^2 + \frac{dV_{\rm{int}}}{d \left|\Phi\right|^2}\right)\Phi
~,\label{FB_EKG1}\\
\nabla^{a}T^{M}_{ab} &=& 0~,\label{mattter_FB1}\\
\nabla_{a}(\rho u^{a}) &=& 0~.\label{mattter_FB2}
\end{eqnarray}
The equilibrium configuration equations for a single fermion-boson can be obtained by combining the same techniques to build isolated non-rotating neutron and boson star solutions, respectively. The method can be summarised as follows: assuming a polar static metric (in Schwarzschild-polar coordinates), impose a harmonic time dependence for the scalar field and the static fluid condition, i.e., $v^{i}=0$. Thus, the Einstein-Klein-Gordon-Euler equations become a simple set of ordinary differential equations that can be numerically integrated under suitable regularity and boundary conditions. More details can be found in ~\cite{Liebling:2012fv}.

Fermion-boson stars have been extensively used as a laboratory for dark matter in the context of neutron stars. Here, dark matter particles might form as a Bose-Einstein condensate, which can be represented with a single complex scalar field~\cite{Brito:2015yfh, Brito:2016peb}. Note that in this model, the coupling between the boson and the fermion particles is just through gravity. One of the first numerical simulations was performed by~\cite{Valdez-Alvarado:2012rct} considering a quadratic massive potential. The authors performed dynamical evolution using numerical relativity techniques in spherical symmetry. Due to the coupling between the fermionic and bosonic components through gravity, new oscillation modes were discovered in the fermionic sector. These new modes are significant as they provide evidence of dark matter's presence within these compact objects. Lastly, a quartic self-potential has been explored in \cite{Valdez-Alvarado:2020vqa}. Dynamical formation and stability of fermion-boson stars have been recently studied in Ref.~\cite{DiGiovanni:2020frc}. The authors showed that an existing neutron star surrounded by an accreting cloud of scalar particles settles down, through gravitational cooling, dynamically into an equilibrium-stable configuration asymptotically consistent with a static solution of the Einstein-Klein-Gordon-Euler system. A follow-up of non-linear stability was conducted in~\cite{DiGiovanni:2021vlu}, finding via numerical-relativity simulations a collaborative stabilisation mechanism whereby a sufficient amount of fermionic matter can stabilise an unstable excited scalar field. Interestingly, fermion-boson stars could help explain recent multi-messenger observations of compact stars~\cite{DiGiovanni:2021ejn}.

The fermionic part could also be modelled by a white dwarf, with different observational consequences~\cite{Leung:2019ctw,Sanchis-Gual:2022ooi,Chan:2023atg}. In~\cite{Sanchis-Gual:2022ooi}, the authors explored the dynamical formation of a spherical mixed white-dwarf--boson star undergoing gravitational cooling for the first time, finding new features in various aspects of the white dwarf, including its gravitational redshift and expected electromagnetic radiation emitted from its photosphere. Additionally, differentially rotating neutron stars experiencing a bar-mode instability have been explored through numerical relativity simulations in the context of fermion-boson stars, as discussed in~\cite{DiGiovanni:2022mkn}. This study found that the accretion of dark matter onto the neutron star could alter the frequency of gravitational-wave emission expected from the bar-mode instability, thereby having an intrinsic impact on the search for continuous gravitational waves. 

It should be mentioned that although most previous works have only included quadratic $\Phi^2$  or quartic $\Phi^4$ potentials, in Ref.~\cite{DiGiovanni:2022xcc}, the authors considered a function potential that models axions and studied static solutions and non-linear stability through simulations in spherical symmetry. In the context of neutron star and dark matter gravitation-only interaction, some progress has been made in understanding non-linear evolutions in spherical symmetry. Research has been conducted on a two-fluid approach in~\cite{Gleason:2022eeg} and a fermion-boson star with a realistic equation of state in~\cite{Nyhan:2022pda}.

\begin{figure}[h!]
\centering
\includegraphics[width=0.5\linewidth]{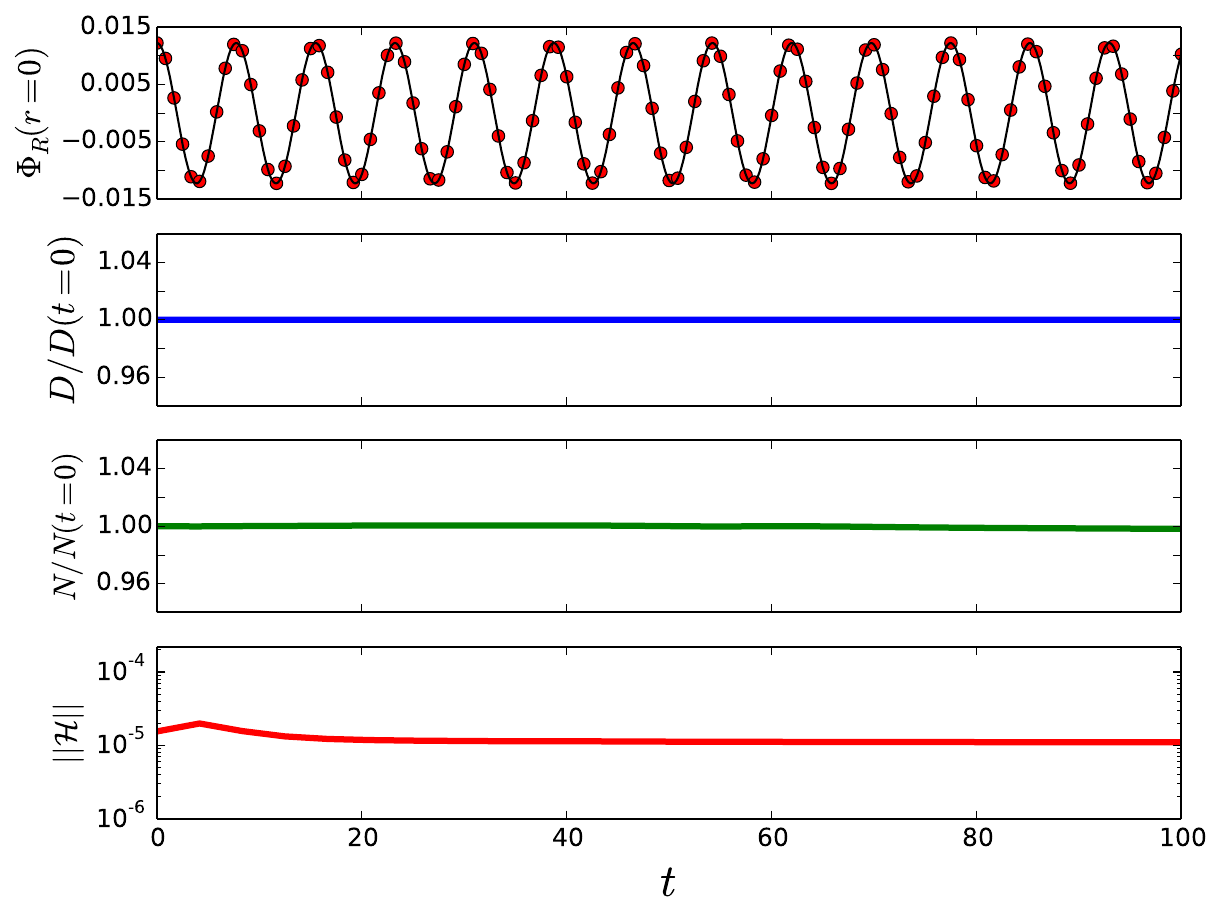}\hspace{-0.01\linewidth}
\includegraphics[width=0.5\linewidth]{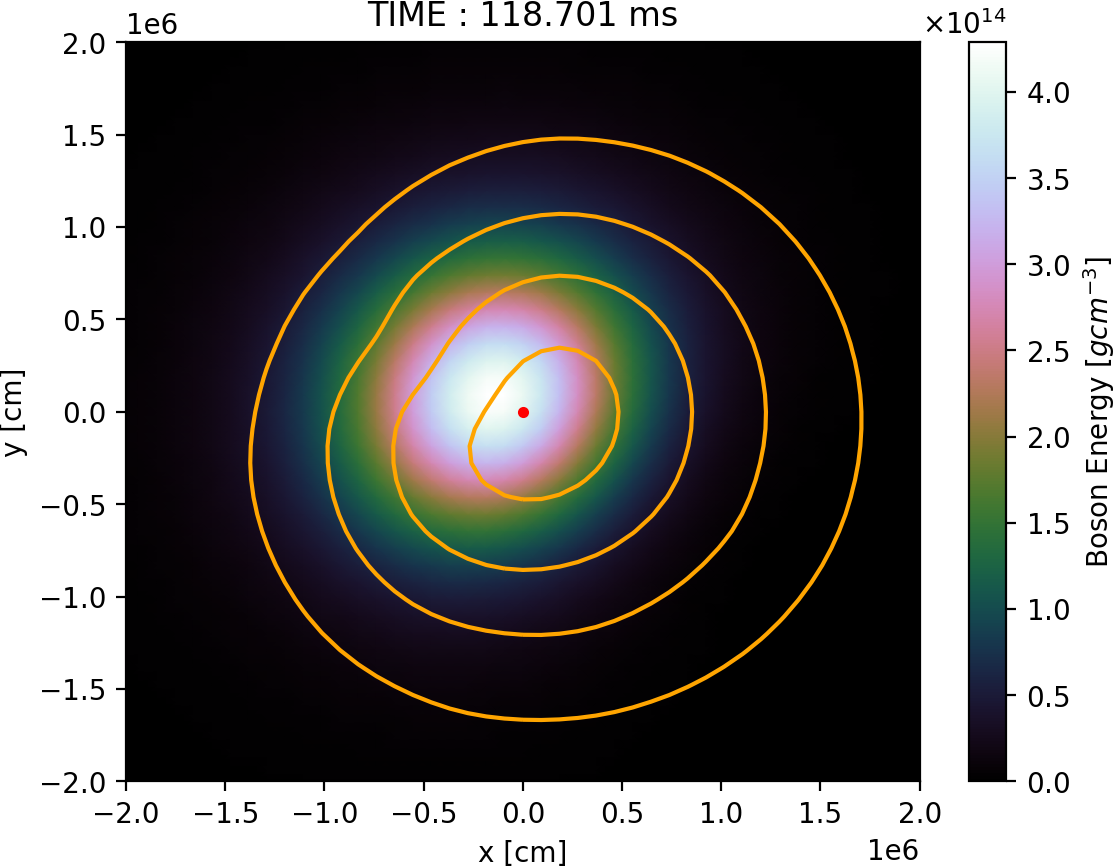}\hspace{0.02\linewidth}\\
\caption{ {\em Fermion-Boson star.} The left panel shows the evolution of an isolated fermion-boson star. (Left panel) Evolution of the real part of $\Phi$ at $r=0$ for a stable configuration. (Second and third row) Evolution of the integrated rest-mass density $D$ and the Noether charge $N$. (Right panel)  $L_{2}$ norm of the Hamiltonian $\mathcal{H}$ as a function of time. Extracted rom~\cite{Bezares:2019jcb}. The right panel displays the morphology of a mixed-bar where the color represent the bosonic energy density and the isocontours represent the fermionic rest-mass density on the equatorial plane. Extracted from~\cite{DiGiovanni:2022mkn}.}
\label{hamandbar}
\end{figure}

Fig.~\ref{hamandbar} shows the dynamically stable evolution of an isolated fermion-boson star configuration and the morphology of a \textit{mixed-bar} made of ultralight bosonic field dark matter and fermionic matter on the dynamics of unstable differentially rotating neutron{-boson} stars.

\subsection{Anisotropic stars and fuzzballs}

Anisotropic stars are compact objects introduced to study mainly two different physical aspects: the first one is in the context of rescuing some equation of state that has been ruled out by astrophysical observations~\cite{Baiotti:2019sew}, and the second regarding Buchdahl’s theorem~\cite{Buchdahl:1959zz} and ultra-compact objects that can violate this theorem~\cite{Cardoso:2019rvt}. Here, we will focus on the latter; for a complete set of references about the former topic, see the bibliography therein in ~\cite{Baiotti:2019sew}. Buchdahl's theorem states the maximum compactness of a self-gravitating, isotropic, spherically-symmetric object of mass $M$ and radius $R$ is $M/R= 4/9,$ if the object is composed of a perfect fluid. As a counter-example, Ref.~\cite{Raposo:2018rjn} presents the first fully covariant and self-consistent model for spherically symmetric anisotropic fluids in General Relativity. The anisotropy was imposed using a relation through the tangential pressure and the radial one, through a parameter $\mathcal{C}$, which measures the deviation between both pressures. This model allows for the existence of stable solutions, some of which can be extremely compact, approaching $M/R = 1/2$ for different parameter values of $\mathcal{C}$. In spherical symmetry, the non-linear evolution of these anisotropic stars was performed for the first time in Ref.~\cite{Raposo:2018rjn}. The authors found that the linearised stability analysis is in suitable agreement with the fully non-linear evolutions for these stars, showing that such configurations are stable below the maximum mass.

Another exotic compact object emerges when we want to evade the Buchdahl limit~\cite{PhysRev.116.1027}, the so-called fuzzball~\cite{Lunin:2001jy,Lunin:2002qf,Mathur:2005zp,Mathur:2008nj}, which can be defined as a microstate geometry with the exact mass, charge, and angular momentum of a black hole, but differing in the structure of the horizon. This object is a horizonless compact object driven by string theory with a rich multipole structure~\cite{Bianchi:2020bxa}. Their echoes have been computed numerically~\cite{Bena:2020see}. Non-linear evolutions have been performed in Ref.~\cite{Ikeda:2021uvc}, where the authors carried out numerical simulations of small fields propagating on a large class of microstate geometries for the first time, focusing on the entire ringdown phenomenology. More details on these exotic compact objects can be found in the following reviews~\cite{Cardoso:2019rvt,Maggio:2021ans}.

\begin{figure}[h!]
\centering
\includegraphics[width=0.5\linewidth]{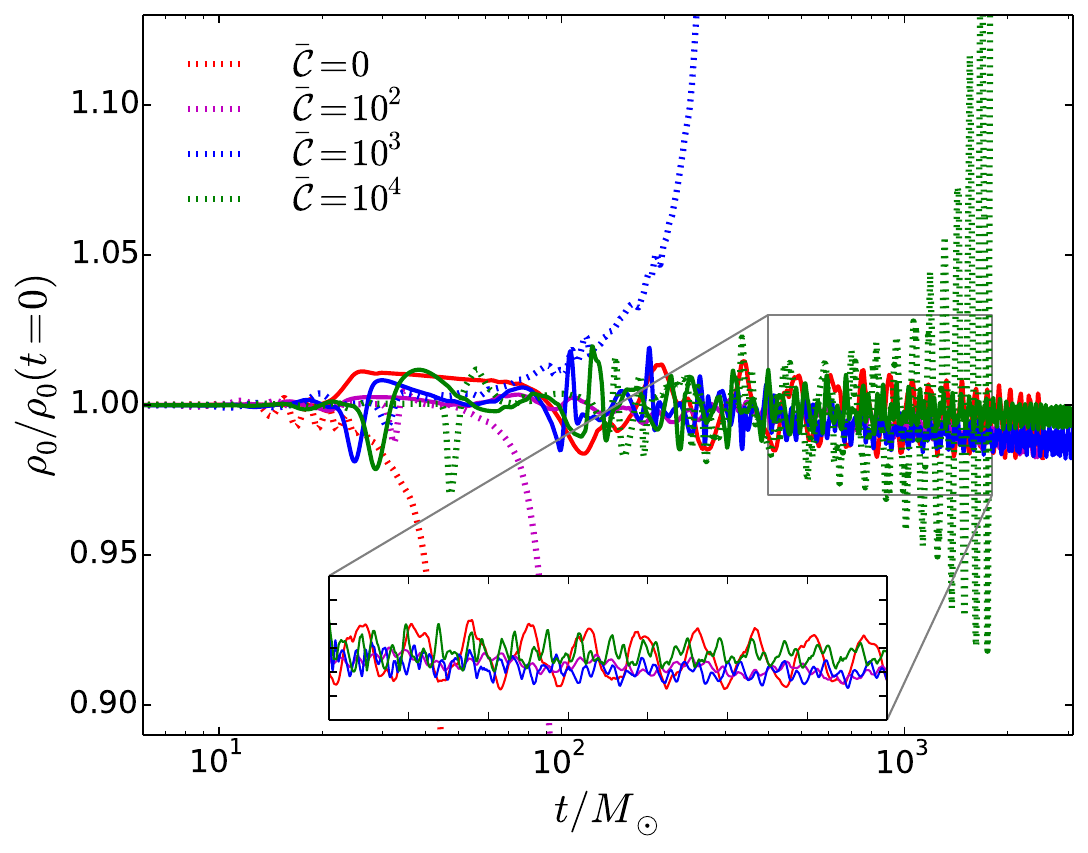}\hspace{-0.01\linewidth}
\includegraphics[width=0.45\linewidth]{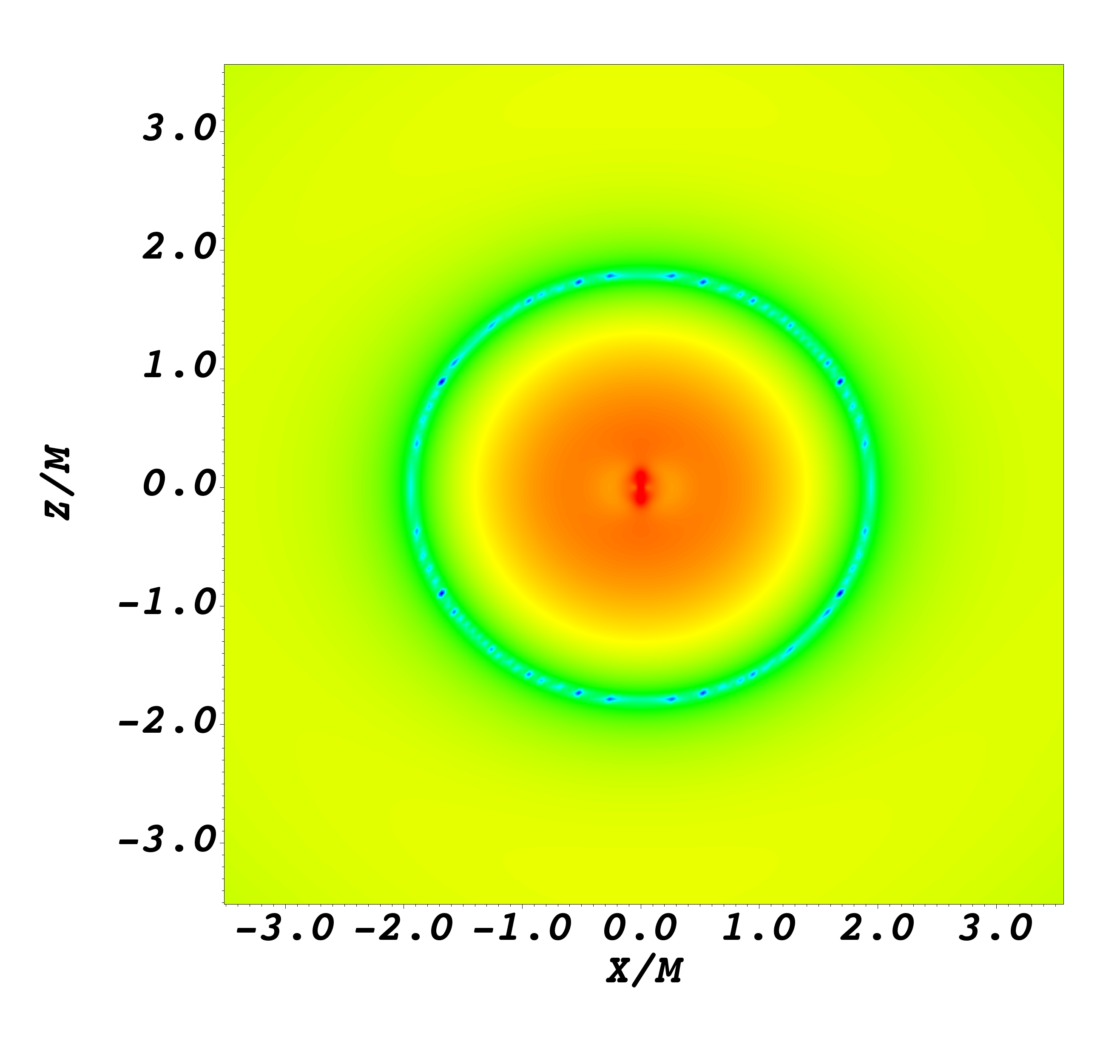}\hspace{0.02\linewidth}\\
\caption{ {\em Anisotropic stars and fuzzball.} The left panel exhibits the normalised central density evolution for different stable(continuous curves)/unstable(dotted curves) configurations. Extracted from~\cite{Raposo:2018rjn}. The right panel shows a snapshot of the evolution of a scalar field around an axisymmetric fuzzball. Extracted from~\cite{Ikeda:2021uvc}.}
\label{anifuzz}
\end{figure}

Finally, Fig.\ref{anifuzz} displays representative plots from the evolution of anisotropic neutron stars and fuzzball taken from~\cite{Raposo:2018rjn,Ikeda:2021uvc}.

\subsection{Strange quark stars}~\label{strange}

The gravitational-wave event GW170817~\cite{LIGOScientific:2017vwq} opened a new window into the interior of neutron stars and the potential identification of the cold equation of state of dense matter~\cite{LIGOScientific:2018cki}, corresponding to densities above $\sim10^{15}$ g/cm$^{3}$, describing these objects. The different equations of state proposals from nuclear physics result in distinct properties~\cite{compose,Ozel:2016oaf}, such as compactness, maximum mass, and radii, that can now be tested with multi-messenger observations, ruling out some of them. In this regard, an interesting hypothesis is that strange quark matter is the true ground state of matter~\cite{Witten:1984rs}. These exotic stars could be composed of pure strange quarks or even undergo phase transitions to quark matter during the merger phase, with neutrons and quarks coexisting in dense matter~\cite{Chakrabarty:1991ui,Buballa:2014jta,Bauswein:2018bma}. Rotating solutions have also been obtained in~\cite{Lin:2006sm,Zhou:2017xhf,Zhou:2019hyy}, both with uniform and differential rotation. Understanding the dynamics and stability of hybrid and strange quark stars is a necessary task to identify the viability of these models and the potential imprints of quark matter and phase transition in their gravitational wave emission. Moreover, an additional remarkable feature of strange stars is that they possess a crust, which is a discontinuity in the density between its surface and the exterior. A typical equation of state is given by the MIT bag model~\cite{Chodos:1974je,Chodos:1974pn}:
\begin{equation}
    P=\frac{1}{3}(e-4B),
\end{equation}
where $P$ is the pressure, $e$ the energy density, and $B$ the bag constant. The pressure goes to zero at a non-zero density, which presents a large jump at the surface. 

It is this sharp transition at the stellar surface that makes the numerical study of these objects a challenge. Numerical codes solving the general relativistic hydrodynamic equations often struggle with large gradients in the hydrodynamic variables (density, pressure). For instance, shocks are dealt with High-Resolution Shock Capturing schemes~\cite{Marti:1991wi,romero1996new}. The idea behind such schemes is to solve local Riemann problems, making use of the hyperbolicity of the hydrodynamic equations. Discontinuities can cause many issues in numerical simulations, therefore new techniques were needed to be able to evolve even single isolated strange stars.

Numerical evolutions have been reported in ~\cite{Zhu:2021xlu, Zhou:2021upu, Zhou:2021tgo,Chen:2023bxx}. In ~\cite{Zhu:2021xlu}, a novel approach to tackle the modelling of large discontinuities at the surface was described. This technique consists of attaching a thin hadronic ``crust'' described by a polytropic equation of state to the quark star, with the density going smoothly to zero. The addition of the crust only changes slightly (around $\sim1\%$) the tidal deformability, according to the authors, and it is shown to allow the correct handling of the stellar surface. They also found that, in general, quark stars seem to require more resolution due to the sharp surface than a hadronic star of similar parameters. Binary strange star mergers are also investigated, and this will be discussed in the following Section~\ref{bnstrs}. 

Another approach implemented in~\cite{Zhou:2021upu} to evolve bare quark stars without a thin nuclear matter crust consists of modifying the recovery primitive procedure (recovering the primitive hydrodynamic variables from the conserved ones) to handle the surface discontinuity and the specific enthalpy going to unity when the pressure goes to zero. The authors successfully performed time evolutions of two triaxially deformed and differentially rotating quark star models with different masses and analyze their gravitational wave emissions. Addressing the single star and binary evolutions of triaxial quark stars is relevant since supramassive models (with masses larger than the maximum value coming from the Tolman-Oppenheimer-Volkoff equations) exist in larger parameter space than neutron stars, making them a potential outcome of supernova explosions or binary mergers. However, the evolution of isolated strange stars is found to be qualitatively similar to that of neutron stars, with a comparable gravitational-wave emission. Besides, for supramassive configurations, the emission decays faster as it settles quickly to axisymmetry. Therefore these models, and in particular the supramassive ones, are not necessarily easier to identify since they are not a better source for continuous gravitational waves. On the other hand, differentially rotating strange stars can still suffer bar-mode instability in a similar timescale than neutron stars.

Finally, in Ref.~\cite{Chen:2023bxx}, stable evolutions of rotating bare quark stars close to the Keplerian limit were performed. In this case, another numerical technique is employed. The surface discontinuity is treated, making use of high-resolution shock-capturing methods together with a Riemann solver that preserves the positivity of density and pressure and a dust-like atmosphere. This allows them to study the oscillating modes and the universal relations of quark stars.

\subsection{Black holes with synchronised hair and superradiance}

An astounding property of black holes is that they can be completely characterized by a set of three parameters: mass, angular momentum, and electric charge. In the non-charged case, from the stellar to the supermassive mass ranges, all these objects are fundamentally described by the same solution (Kerr hypothesis). This is the no-hair theorem, preventing the existence of new black hole solutions in General Relativity. However, uniqueness theorems can be circumvented if we relax some of the assumptions, for instance, considering a non-electrovacuum scenario taking a complex and massive bosonic field around a Kerr black hole. Here, we will present some examples of hairy black holes in General Relativity with minimally coupled massive scalar and vector fields~\cite{Herdeiro:2014goa,Herdeiro:2015waa,Herdeiro:2016tmi}.

Before discussing these hairy black hole solutions, let us note that a bosonic field scattered off a rotating black hole can be amplified due to the superradiance process~\cite{Brito:2015oca} and extract energy and angular momentum from the black hole if the oscillation frequency of the bosonic field, $\omega$, satisfies the following condition:
\begin{equation}
    \omega<m\Omega_{\rm{H}},
\end{equation}
where $m$ is the azimuthal winding number and $\Omega_{\rm{H}}$ is the angular velocity at the horizon. This phenomenon occurs for scalar, vector, and tensor fields, including electromagnetic and gravitational waves~\cite{East:2013mfa}. Therefore, if these superradiant-scattered waves can be trapped around a black hole by means of a cavity or a potential well due to the mass term (for massive bosonic fields), the superradiant instability can be triggered. Superradiant bosonic modes will grow exponentially and continuously extract energy in an unstoppable process leading to the black hole bomb scenario~\cite{Press:1972zz} when the field reaches the non-linear regime. Such a system was investigated in linear studies~\cite{Cardoso:2004nk,Dolan:2012yt,Herdeiro:2013pia}. However, the endpoint of such instability remained an open question due to the long timescales involved~\cite{Dolan:2012yt,Okawa:2014nda}. Consequently, fully non-linear numerical-relativity simulations were needed to answer this question.

A most favourable scenario for numerical simulations involving {\it charged} black holes instead of rotating solutions were explored in 2016. Reissner-Nordstr\"om black holes can suffer an analogue process to superradiant scattering by a charged bosonic field that extracts Coulomb energy and electric charge given rise to the charged black hole bomb~\cite{Bekenstein:1973mi}. In this case the superradiant condition relates $\omega$ with the electric potential at the horizon $\phi_{\rm H}$ and the charge parameter of the bosonic field $q$:
\begin{equation}\label{supcharge}
\omega < q \phi_{\rm H}.    
\end{equation}
This problem can be tackled by solving the Einstein-(charged, complex) Klein-Gordon-Maxwell system in spherical symmetry. Moreover, the growth rates of the instability are orders of magnitude smaller than the rotating counterpart~\cite{Herdeiro:2013pia}, significantly reducing the computational cost. A trapping mechanism must be included by imposing reflecting boundary conditions since the mass term is not enough to counter the electric repulsion between the charged black hole and the charged bosonic field. Numerical simulations showed that the development of the superradiant instability leads not to an explosive phenomenon but instead to the dynamical formation of a new static configuration at the threshold of the instability: a charged hairy black hole~\cite{Ponglertsakul:2016wae,Sanchis-Gual:2015lje,Bosch:2016vcp,Sanchis-Gual:2016tcm} (see also \cite{Baake:2016oku}). This process consists of two phases:
\begin{svgraybox}
\begin{enumerate}
    \item while $\omega < q\phi_{\rm H}$: confined superradiant modes extract energy and charge from the black hole, and therefore its electric potential, $\phi_{\rm H}$, decreases;
    \item until we reach $\omega = q\phi_{\rm H}$: the instability saturates and the extraction stops, the bosonic field is now in equilibrium with the black hole forming a new static solution.
\end{enumerate}
\end{svgraybox}

\begin{figure}[h]
\sidecaption
\includegraphics[width=0.62\linewidth]{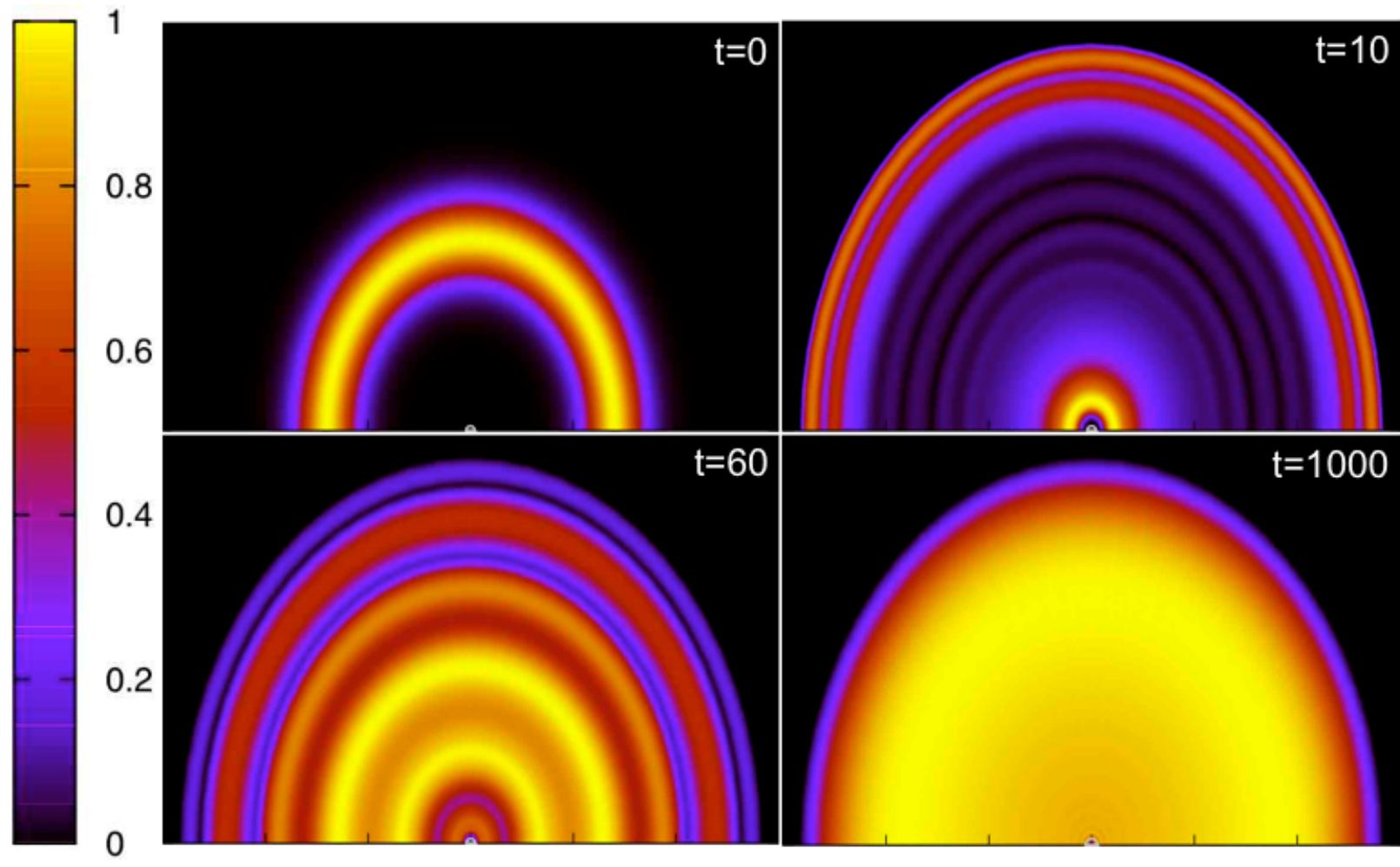}
\caption{ {\em Black holes with synchronised hair and superradiance.} 2D
snapshots of the normalized scalar field radial profile for a scalar charge $qM=40$ at
times t/M = 0, 10, 60, 1000. The apparent horizon is represented by small white circles near the origin in the 2D panels. Extracted from~\cite{Sanchis-Gual:2015lje}.}
\label{fig:125_sup}       
\end{figure}

This evolution is shown in Figure~\ref{fig:125_sup}. The first three panels correspond to the exponential growth of the scalar field energy, whereas the last one displays the final state of the charged hairy black hole.

While the charged case is considered to be only a toy model since astrophysical black holes are not expected to be electrically charged and an artificial cavity must be placed, in 2017, pioneering numerical studies in the fully non-linear regime showed that a qualitatively identical evolution of the superradiant instability of a Proca field was also found in the Kerr case~\cite{East:2017ovw,East:2017mrj}. They also observed the exponential growth and saturation phases, replacing the extraction of electric charge with the extraction of angular momentum. Considering Proca fields instead of scalar fields was motivated by the timescale of the superradiant instability in the scalar case ($t/M\sim10^{7}$), where the current numerical relativity techniques and computational power are still insufficient to perform such simulations. However, Proca fields present shorter timescales of the order of $t/M\sim10^4-10^{5}$ (still a computational prowess!). In order to evolve Proca fields using the standard 3+1 split, the following 3+1 quantities are used:
\begin{eqnarray}
\mathcal{A}_{a} =&\mathcal{X}_{a}+n_{a}\mathcal{X}_{\phi},~~\mathcal{X}_{i} = \gamma^{a}_{\,i}\mathcal{A}_{a},~~\mathcal{X}_{\phi} =-n^{a}\mathcal{A}_{a},
\end{eqnarray}
where $n^{a}$ is the timelike unit vector, $\gamma^{a}_{b}=\delta^{a}_{b}+n^{a}n_{b}$ is the operator projecting spacetime quantities onto the spatial hypersurfaces, $\mathcal{X}_{i}$ is the vector potential, and $\mathcal{X}_{\phi}$ is the scalar potential. The Proca evolution equations are then given by~\cite{Zilhao:2015tya}:
\begin{eqnarray}
\label{eq:dtAi}
\partial_{t} \mathcal{X}_{i}      & = & - \alpha \left( E_{i} + D_{i} \mathcal{X}_\Phi \right) - \mathcal{X}_\Phi D_{i}\alpha + \mathcal{L}_{\beta} \mathcal{X}_{i}
,\\
\label{eq:dtE}
\partial_{t} E^{i}       & = &
        \alpha \left( K E^{i} + D^{i} Z + \mu^2\mathcal{X}^{i}
                + \epsilon^{ijk} D_{j} B_{k} \right)
        - \epsilon^{ijk} B_{j} D_{k}\alpha
        + \mathcal{L}_{\beta} E^{i},\\
\label{eq:dtAphi}
\partial_{t} \mathcal{X}_\Phi  & = & - \mathcal{X}^{i} D_{i} \alpha
        + \alpha \left( K \mathcal{X}_\Phi - D_{i} \mathcal{X}^{i} - Z \right)
        + \mathcal{L}_{\beta} \mathcal{X}_\Phi ,\\
\label{eq:dtZ}
\partial_{t} Z          & = & \alpha \left( D_{i} E^{i} + \mu^{2} \mathcal{X}_\Phi - \kappa Z \right)
        + \mathcal{L}_{\beta} Z\,,
\end{eqnarray}
with $Z$ an additional variable and $\kappa$ a damping parameter that help stabilize the numerical evolution, $\epsilon^{ijk}$ the Levi-Civita symbol, and the three-dimensional ``electric" $E^{i}$ and ``magnetic" $B^{i}$ fields, in analogy with Maxwell theory, are defined as
\begin{equation}
E_{i}=\gamma^{a}_{i}\mathcal{F}_{ab}n^{b},\quad B_{i}=\gamma^{a}_{i}\star\mathcal{F}_{ab}n^{b}=\epsilon^{ijk}D_{i}\mathcal{X}_{k}.
\end{equation}
The stress-energy tensor is given by:
\begin{eqnarray}
    T_{ab}&=&\frac{1}{2}(\mathcal{F}_{ab}\mathcal{F}^{*}_{bd}+\mathcal{F}^{*}_{ag}\mathcal{F}_{bd})g^{gd}-\frac{1}{4}g_{ab}\mathcal{F}^{gd }\mathcal{F}^{*}_{g d}
\nonumber \\ &&+\frac{\mu^2}{2}(\mathcal{X}_a\mathcal{X}^{*}_b+\mathcal{X}^{*}_a \mathcal{X}_b-g_{ab}\mathcal{X}^d\mathcal{X}^{*}_d)~.
\end{eqnarray} 
Such simulations showed numerically that the growth of a Proca field perturbation around a Kerr black hole, when reaching the non-linear regime, leads to the formation of a Kerr black hole with synchronised Proca hair at the endpoint of the superradiant instability~\cite{Herdeiro:2017phl}. The potential formation of Kerr black holes with synchronised bosonic hair was also found in the context of boson and Proca mergers, which leads to a final bosonic cloud with a larger $m\gtrsim 4-6$ depending on the progenitors~\cite{Sanchis-Gual:2020mzb,Bezares:2022obu}. Additionally, in 2018, the nonlinear evolution of the superradiant instability of rotating black holes in AdS$_{4}$ considering gravitational waves and solving the fully non-linear 3+1 Einstein equations~\cite{Chesler:2018txn}. 

Furthermore, Proca field with self-interaction has also been explored within numerical relativity. It was found that this theory can lead to significant problems, such as the loss of hyperbolicity of the field equations~\cite{Coates:2022nif,Barausse:2022rvg} or the generation of instabilities in the context of superradiance~\cite{Clough:2022ygm}.

\section{Beyond General Relativity theory}\label{Section2}

General Relativity faces numerous conceptual problems, such as singularity formation, horizons and the information loss paradox, its breakdown as a quantum theory at energies that exceed the Planck scale, which indicates the need for a new theory or a non-perturbative approach, or the nature of dark matter and dark energy, that seems to indicate that it is not the ultimate theory of gravity. Modified gravity theories have attempted to solve all or a least some of these issues. Within these extended theories, new aspects of well-known compact objects or even new exotic compact objects have been found.

\begin{svgraybox} In this section, we will discuss the numerical simulations of boson stars in modified theories of gravity and the impact of an additional scalar degree of freedom on black holes and neutron stars.\end{svgraybox}

\subsection{Boson stars in modified gravity}

Bosonic stars can also be studied within new theories of gravity beyond General Relativity as new compact objects with different properties or to explore new phenomena that arise in these theories, for example, related to spontaneous scalarization~\cite{Damour:1996ke} or during gravitational collapse and the formation of black holes. Equilibrium solutions have been found in a number of theories, such as 
\begin{itemize}
    \item scalar-tensor gravity~\cite{Comer:1997ns,Whinnett:1999ma,Herdeiro:2018wvd,Brihaye:2019puo}, 
    \item Brans-Dicke theory~\cite{Torres:1998xw},
    \item Einstein-Dilaton-Gauss-Bonnet gravity~\cite{Pani:2011xm}, 
    \item Einstein-Gauss-Bonnet theory~\cite{Hartmann:2013tca}, 
    \item teleparallel gravity~\cite{Horvat:2014xwa}, 
    \item higher-derivative gravity~\cite{Baibhav:2016fot},
    \item Horndenski gravity~\cite{Brihaye:2016lin,Roque:2021lvr}
    \item tensor-multi-scalar theories~\cite{Yazadjiev:2019oul}, 
    \item $f(T)$ extended theory~\cite{Ilijic:2020vzu}, or
    \item Palatini $f(R)$ theory~\cite{Maso-Ferrando:2021ngp}. 
\end{itemize}

Despite the attention received, not many numerical simulations of boson stars in modified gravity have been carried out, generally due to the lack of well-tested evolution formalisms. Simulations have been reported in at least scalar-tensor ~\cite{Whinnett:1999ma,Alcubierre:2010ea}, Brans-Dicke~\cite{Balakrishna:1997ek} and more recently in Palatini $f(R)$ theories~\cite{Maso-Ferrando:2023nju,Maso-Ferrando:2023wtz}. The evolutions are usually performed in the so-called Einstein frame given by a conformal transformation of the metric in the physical ``Jordan" frame since the former delivers equations that are similar to those in General Relativity and where the non-linearities in the gravity sector can be transferred to the matter Lagrangian. The relevant quantities can then be easily recovered by means of conformal transformations. In~\cite{Maso-Ferrando:2023nju,Maso-Ferrando:2023wtz}, the simulations were also done in the Einstein frame instead of the $f(R)$ frame, but a mapping is needed to recover and study the true dynamics in Palatini $f(R)$~\cite{Afonso:2018bpv}.

The action of a boson star in Palatini $f(R)=R+\xi R^2$ gravity, with $\xi$ a parameter of the theory, in the Einstein frame  is~\cite{Maso-Ferrando:2023nju,Maso-Ferrando:2023wtz}
\begin{equation}\label{eq:ac2}
			S_{\rm EF}=\int d^4x \sqrt{-q} \frac{R}{2 \kappa}-\frac{1}{2}\int d^4x \sqrt{-q} \left(\frac{Z-\xi \kappa Z^2-2V(\Phi)}{1-8 \xi \kappa V(\Phi)} \right) \quad ,
	\end{equation}
where the kinetic term $Z\equiv q^{ab}\partial_{a}\Phi^*\partial_{b}\Phi$ is now contracted with the (inverse) metric $q^{ab}$, $V(\Phi)=-\mu^2 \Phi^* \Phi/2$, $\kappa=8\pi$ (in $G=c=1$ units), and $R$ is the Ricci scalar of the metric  $q_{ab}$, i.e., $R= q^{ab}R_{ab}(q)$. In spherical symmetry, the space-time metric in the Einstein frame is given by
\begin{equation}\label{eq:EinsteinMetric}
    \begin{aligned}
        ds_{\text{EF}}^2=&-(\alpha^2 -\beta^x \beta_{x})dt^2+2\beta_{x}dx dt \\
        &+ e^{4 \chi(t,x)}\left( a(t,x)dx^2+x^2 b(t,x) d\Omega^2\right)\,\,,\\
    \end{aligned}
\end{equation}
where $x$ is a radial coordinate, $d\Omega^{2}=d\theta^2+\sin^2\theta d\varphi^2$, $a(t,x)$ and $b(t,x)$ are the conformal metric components, and $\chi(t,x)$ is a conformal factor.

Two first-order variables are introduced to reformulate the Klein-Gordon equation 
	\begin{eqnarray}
		\Psi  := \partial_{x}\Phi\,\, ,~~\Pi := n^{\mu}\partial_{\mu}\Phi 
			=\frac{1}{\alpha}\left( \partial_{t}\Phi-\beta^{x}\Psi\right)\,\, .
	\end{eqnarray}
 Therefore, the equations of motion for the scalar field are
	\begin{eqnarray}
		\partial_{t}\Phi&=&\beta^{x}\partial_{x}\Phi+\alpha \Pi\,\,,\\
		\partial_{t}\Psi&=&\beta^{x}\partial_{x}\Psi+\Psi\partial_{x}\beta^x+\partial_{x}\left(\alpha \Pi\right)\,\,,\\
		\partial_{t}\Pi&=&\frac{1-2\kappa \xi Z+\kappa \xi |\Pi|^2}{1-2\kappa \xi Z+2\kappa \xi |\Pi|^2}\left\{\Xi-\frac{\kappa \xi\Pi^2 \Xi^*}{1-2\kappa \xi Z+\kappa \xi |\Pi|^2}\right\}\,\,,
  \nonumber \\
	\end{eqnarray}
where the new variable $\Xi$ simplifies the notation: 
	\begin{align}
        \Xi:=&\beta^{x}\partial_{x}\Pi+\frac{\Psi}{ae^{4\chi}}\partial_{x}\alpha+\frac{\alpha}{a e^{4\chi}}\left[\partial_{x}\Psi+\Psi\left(\frac{2}{x}-\frac{\partial_{x}a}{2a}+\frac{\partial_{r}b}{b}+2\partial_{x}\chi\right) \right]+\alpha K \Pi \nonumber \\
        &-\frac{\alpha \mu^2 \Phi}{1-2\kappa \xi Z}  +\frac{\alpha \left(Z-\kappa \xi Z^2+\mu^2|\Phi|^2\right)4\xi\kappa\Phi\mu^2}{\left(1+4\kappa \xi \mu^2 |\Phi|^2\right)\left(1-2\kappa \xi Z\right)} \nonumber \\
        &-\frac{4\kappa \xi \mu^2 \alpha}{1+4\kappa \xi \mu^2 |\Phi|^2}\left[-\frac{\Pi}{\alpha}\left(\partial_{t}\Phi^*\Phi+\Phi^*\partial_{t}\Phi\right)+\left(\frac{\Psi}{e^{4\chi}a}+\frac{\Pi\beta^x}{\alpha}\right)\left(\partial_{x}\Phi^*\Phi+\Phi^*\partial_{x}\Phi\right)\right]\nonumber \\
        &-\frac{\alpha \kappa \xi}{1-2\kappa \xi Z}\left[-\frac{\left(\partial_{t}\Psi^*\Psi+\Psi^*\partial_{t}\Psi\right)e^{4\chi}a-|\Psi|^2\left(4e^{4\chi}a\partial_{t}\chi+e^{4\chi}\partial_{t}a\right)}{e^{8\chi}a^2}\frac{\Pi}{\alpha}\right.\nonumber \\
        &\left.-\frac{\left(\partial_{x}\Psi^*\Psi+\Psi^*\partial_{x}\Psi\right)e^{4\chi}a-|\Psi|^2\left(4e^{4\chi}a\partial_{x}\chi+e^{4\chi}\partial_{x}a\right)}{e^{8\chi}a^2}\left(\frac{\Psi}{e^{4\chi}a}+\frac{\Pi\beta^x}{\alpha}\right)\right.\nonumber \\
        &\left.-\left(\frac{\Psi}{e^{4\chi}a}+\frac{\Pi\beta^x}{\alpha}\right)\left(\partial_{x}\Pi^*\Pi+\Pi^*\partial_{x}\Pi\right)\right] \,.
		\end{align} 

The main findings establish that boson stars in modified theories behave similarly to those in General Relativity in terms of stability and dynamical formation. Stable solutions oscillate around the initial configuration when perturbed, while unstable ones suffer one of the three expected outcomes: dispersion, migration to the stable branch, or collapse to black hole. Moreover, it has also been shown that boson stars can form in such theories much like they do in General Relativity, through the well-known gravitational cooling mechanism~\cite{Seidel:1993zk}.

\begin{figure}[h]
\sidecaption
\includegraphics[width=0.48\linewidth]{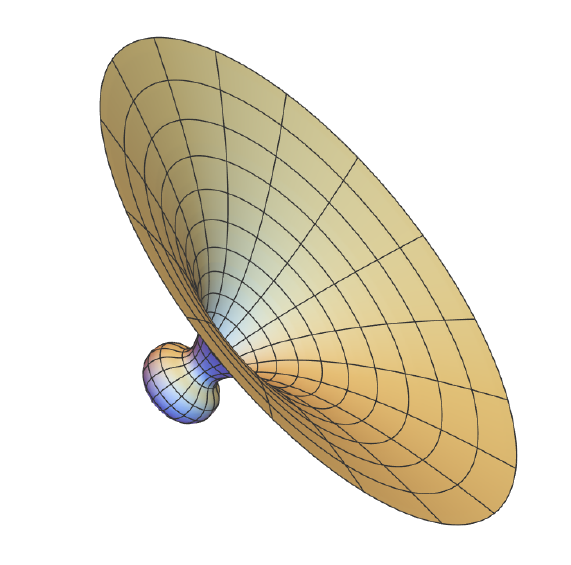}
\caption{ {\em Boson stars in modified gravity.} Embedding diagram of the late-time spacetime geometry from the gravitational collapse of a boson star at two different times. The formation of a throat and a baby universe inside the horizon can be observed. Extracted from~\cite{Maso-Ferrando:2023wtz}.}
\label{fig:131_emb}       
\end{figure}

The gravitational collapse in the particular case of Palatini $f(R)$ theory was studied in~\cite{Maso-Ferrando:2023nju, Maso-Ferrando:2023wtz} focusing on black hole formation from unstable boson stars with a non-linear potential in spherical symmetry. It was found that, while in the Einstein frame, the endpoint of the evolution is a Schwarzschild black hole surrounded by a quasi-stationary cloud of the scalar field, when moving to the $f(R)$ frame, the collapse does not lead to the formation of a singularity. Instead, as shown in Figure~\ref{fig:131_emb}, an expanding baby universe emerges, connected to the original universe through a minimal area surface (throat or umbilical cord), supported by scalar field accretion. The throat is, however, hidden inside the horizon at all times, causally disconnecting the baby universe from external observers. This result may have implications for the understanding of singularities and bounce mechanisms in General Relativity and quantum gravity.

\subsection{Black hole and neutron star in scalar-tensor theories}\label{isolatedbgGR}

As explained above, one of the standard forms of modified General Relativity is adding a scalar field that is non-minimally coupled to gravity. The action gives the classic scalar-tensor theory
\begin{equation}
S=\frac{1}{16\pi\,G}\int d^{4}x\sqrt{-g}\left(\phi R - \frac{\omega(\phi)}{\phi}\nabla_{a}\phi\nabla^{a}\phi - V(\phi)\right) + S_{m}(g_{ab},\psi)~,
\label{actionst}
\end{equation}
where $g$ is the determinant of the spacetime metric $g_{ab},$ $R$ is the Ricci scalar of the metric, $\omega(\phi)$ and $V(\phi)$ are functions of $\phi$ that must be defined to establish a specific theory, $S_{m}$ is is the matter action, and $\psi$ denotes the matter fields. As explained previously, this is also known as the Jordan frame, a mathematical representation of a physical theory. Performing a conformal transformation leads to Einstein’s theory with a minimally coupled scalar field, also known as the Einstein frame. We will not discuss details about these frames; more about this discussion can be found in Ref.~\cite{Sotiriou:2007zu}.

With their fundamental properties and extreme conditions, black holes and neutron stars offer a unique environment for studying fundamental fields and exploring potential deviations from General Relativity. The so-called \textit{spontaneous scalarization} is one of the most fascinating mechanisms in this context. This is a tachyonic instability that initiates the growth of a scalar field when the object’s curvature surpasses a specific threshold value. This phenomenon is known to happen for neutron stars~\cite{Damour:1996ke} in scalar-tensor theories, whereas black holes are protected by no-hair theorems~\cite{Hawking:1972qk}. For a complete review of this topic, we refer to the reviews Refs.~\cite{Herdeiro:2015waa,Sotiriou:2015pka,Barack:2018yly,Quiros:2019ktw,Doneva:2022ewd}. 

One of the first numerical relativity simulations of an isolated black hole was performed in Refs.~\cite{Scheel:1994yr,Scheel:1994yn}, where the authors studied the gravitational collapse in Brans-Dicke theory of gravitation (i.e. considering $\omega=\omega_{0}=\rm{constant}$ and $V(\phi)=0$). These studies focused on the emission of scalar waves (monopolar gravitational waves), confirming the no-scalar-hair theorems and proving that the dynamics of black holes in these theories are not the same as in General Relativity. Similar findings were reported in Refs.~\cite{Shibata:1994qd,Harada:1996wt,Kerimo:1998qu,Kerimo:2000gc,Berti:2013gfa}. In addition, numerical relativity simulations of neutron stars collapsing to a stellar-mass black hole have also been done in scalar-tensor theories~\cite{Novak:1999jg,Gerosa:2016fri,Mendes:2016fby,Sperhake:2017itk,Cheong:2018gzn}. Scalarized neutron stars can shed away a large amount of scalar hair during the collapsing process, with these scalar polarization waves becoming smoking guns that could be potentially observed and measured in gravitational waves events~\cite{Rosca-Mead:2019seq,Rosca-Mead:2020ehn,Kuan:2021yih,Kuan:2022oxs}. Furthermore, in Ref.~\cite{Mendes:2021fon}, the authors carried out 1+1 numerical relativity simulations of neutrons that can acquire hair during their evolution. They studied the spectrum of these stellar oscillations, searching for some traces of those oscillations in the post-merger gravitational wave signal from a merger of two neutron stars.

A theory that has gained considerable attention is scalar-Gauss-Bonnet gravity, where the Gauss-Bonnet invariant $\mathcal{G}=R^2-4R_{ab}R^{ab}+R_{abcd}R^{abcd}$ is added to the action coupled to the scalar field as $\alpha\,f(\phi)\mathcal{G}$, where  $\alpha$  is a coupling constant with dimensions of length square and $f(\phi)$ is a function of the scalar field. For example, if  $f(\phi)=\phi$, this theory circumvents the no-hair theorem and provides hairy black holes~\cite{Kanti:1995vq}. In general, if $f'(\phi_{0})=0,$ this theory admits General Relativity solutions, for some constant $\phi_{0}$. Furthermore, theories which do not satisfy $f''(\phi_{0})\mathcal{G}<0$ exhibit the phenomenon of black hole scalarization. For more details about this phenomenon, see Refs.~\cite{Doneva:2017bvd,Antoniou:2017acq,Silva:2018qhn,Doneva:2022ewd}.

One of the main challenges in fully simulating this theory is the general need for the well-posedness of the initial data (Cauchy) problem~\cite{Papallo:2017qvl}. According to the criteria of Hadamard~\cite{Hadamard10030321135}, the Cauchy problem for a system of partial differential equations is well-posed if a unique solution exists that continuously depends on its initial data. Since proving this statement is difficult, one usually attempts to show that the system of partial differential equations is strongly hyperbolic, a sufficient condition for well-posedness under suitable initial and boundary conditions~\cite{Hilditch:2013sba}. 

\begin{figure}[h!]
\centering
\includegraphics[width=0.8\linewidth]{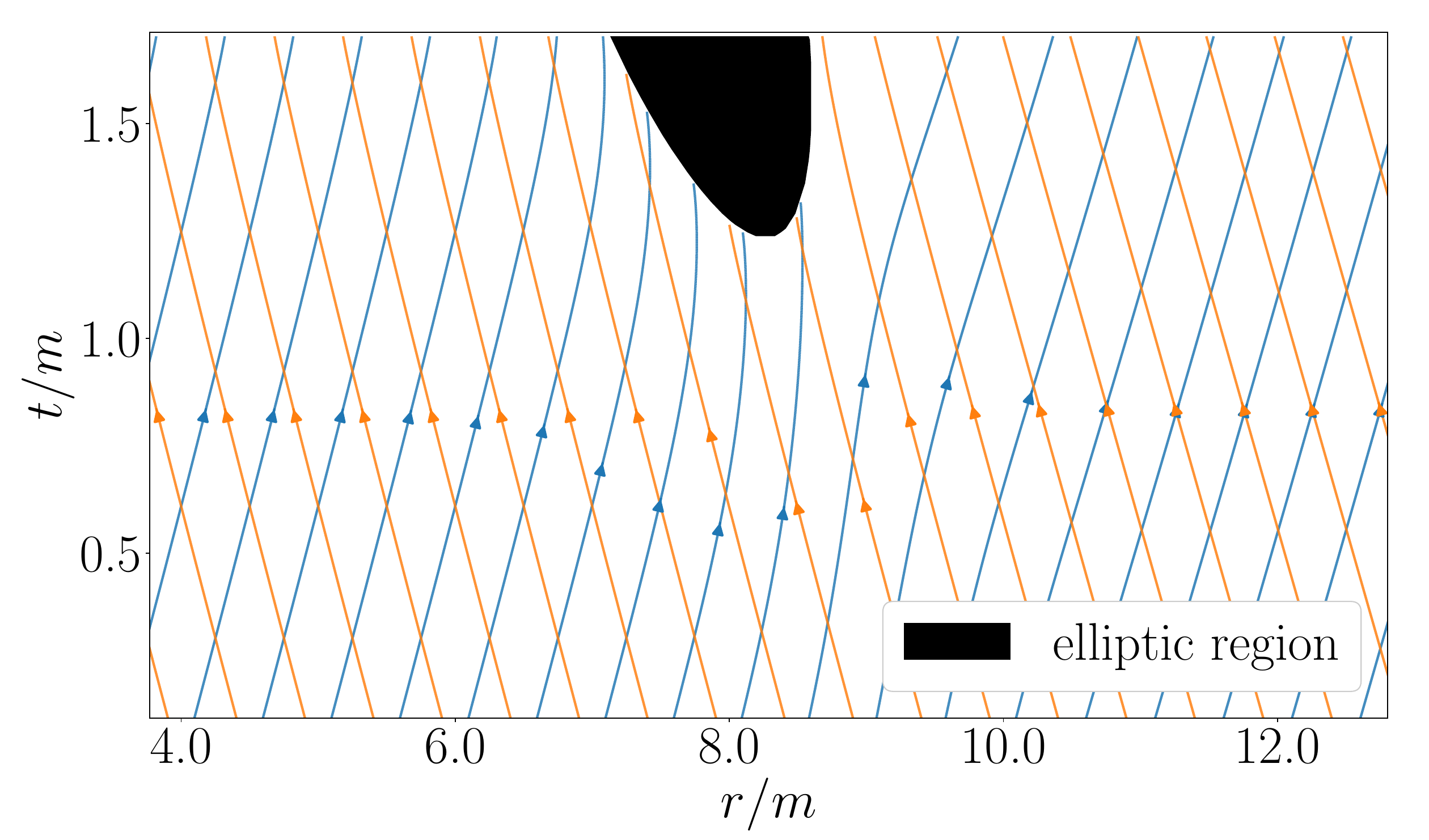}\hspace{-0.01\linewidth}\\
\includegraphics[width=0.85\linewidth]{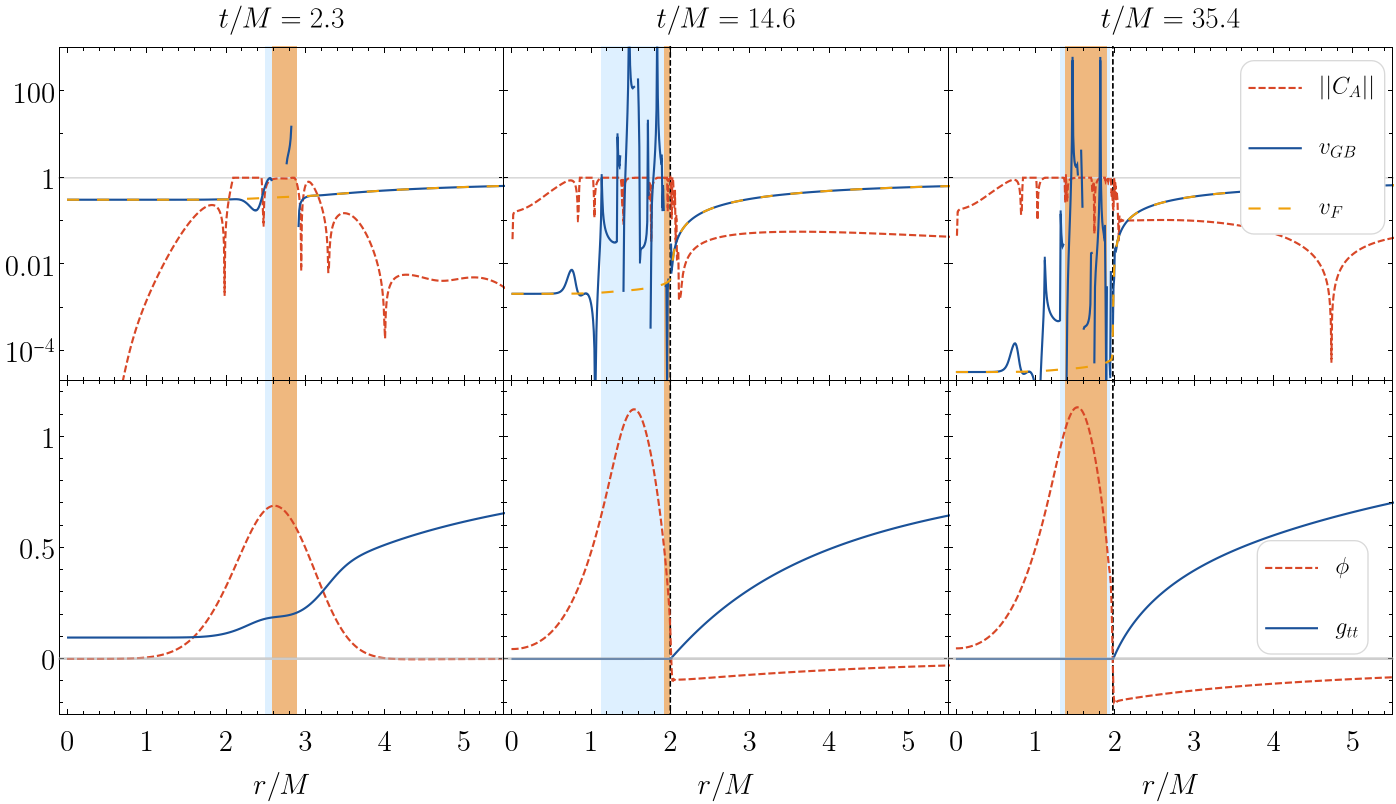}\hspace{0.02\linewidth}\\
\caption{ {\em Black hole and neutron star in scalar-tensor theories.} The top panel shows the elliptic region (i.e., when the characteristic speeds become imaginary, indicating the character of the equations have become elliptic) forming in the simulation of gravitational collapse of  Einstein dilaton Gauss-Bonnet gravity. Extracted from~\cite{Ripley:2019hxt}. The bottom panel displays different time snapshots of various physical quantities, showing how the fixing-the-equation approach allows one to follow the entire simulation until the formation of hairy black holes. Extracted from~\cite{Franchini:2022ukz}.}
\label{bGRsct}
\end{figure}

In Refs.~\cite{Ripley:2019hxt,Ripley:2019irj,Ripley:2020vpk}, the authors studied the gravitational collapse of a scalar field in the shift-symmetric Einstein dilaton-Gauss-Bonnet gravity, see as well~\cite{Corelli:2022phw,Corelli:2022pio}. They found that higher frequency modes develop for specific couplings, leading to the breakdown of the hyperbolic structure of the system. In particular, the authors demonstrated that during the evolution, one of the characteristic speeds becomes imaginary, indicating that the evolution equations switch their character from hyperbolic to elliptic, resembling the mixed-type partial differential equation \textit{Tricomi equation} (see Fig.~\ref{bGRsct}). Furthermore, for sufficiently weak couplings and smooth initial data, local well-posedness of scalar-Gauss-Bonnet gravity can be established~\cite{Kovacs:2020pns,Kovacs:2020ywu} and successful studies have been presented in this regime~\cite{East:2020hgw,East:2021bqk}, which we will discuss in Section~\ref{subsectionbGR}.  

Nevertheless, a successful approach to mitigating the loss of hyperbolicity involves the \textit{fixing-the-equations} technique~\cite{Cayuso:2017iqc,Allwright:2018rut}, inspired by dissipative relativistic hydrodynamics~\cite{1976PhLA...58..213I}. The main idea is to modify the higher-order contributions to the equations of motion \textit{ad hoc} by replacing them with some auxiliary fields and allowing the latter to relax towards their correct values through a driver equation. In Ref.~\cite{Franchini:2022ukz}, in scalar-Gauss-Bonnet theory, the \textit{fixing-the-equations} program has been employed within the gravitational collapse of a scalar field. The study demonstrates that a complete scalarized black hole solution can be achieved after passing a phase during which the original system of field equations changes character, see Fig.~\ref{bGRsct}. This approach has been recently applied in 3+1 simulations in Ref.~\cite{Lara:2024rwa} within the same theory, and it has also been applied in the Effective Field Theory of gravity in Refs.~\cite{Cayuso:2020lca,Cayuso:2023xbc}. Finally, in Ref.~\cite{Thaalba:2023fmq} it was demonstrated that by adding a coupling between the scalar field and the Ricci scalar positively impacts the well-posedness at the nonlinear level. This finding was further confirmed in 3+1 simulations of the evolution of static and rotating black holes~\cite{Doneva:2024ntw}.

Another interesting modified gravity is K-essence, a scalar-tensor theory with first-order derivative self-interactions given by the action 
\begin{equation*}
S=\int \mathrm{d}^{4}x\sqrt{-g}\left[\frac{M_{\mathrm{Pl}}^{2}}{2}R + K(X)  \right]  + S_{m}\left[A(\Phi)g_{ab},\Psi\right]~,
\end{equation*}
where $M_\mathrm{Pl}=(8\pi G)^{-1/2},$ $A(\Phi)$ is the conformal factor and $K(X)$ is a function that depends on the standard kinetic term $X\equiv \nabla_a \varphi \nabla^a \varphi.$ Numerical simulations of gravitational collapse within K-essence, assuming spherical symmetry, have been conducted in~\cite{Akhoury:2011hr,Leonard:2011ce,Bernard:2019fjb,Figueras:2020dzx,Bezares:2020wkn,Lara:2021piy}. These works have shown that there are classes of theories for which the Cauchy problem breaks, encountering again the issue of the change of character from hyperbolic to elliptic (Tricomi)~\cite{Bernard:2019fjb}. Furthermore, in Refs.~\cite{Bernard:2019fjb,Bezares:2020wkn}, it was found that in these theories, there may be divergences of the characteristic speeds, similar to the mixed-type Keldysh equation~\cite{Bernard:2019fjb,Bezares:2017mzk}. Furthermore, k-essence can screen interactions with the scalar field on local scales while allowing for significant deviations from General Relativity on cosmological scales~\cite{Babichev:2013usa}. This screening mechanism is known as k-mouflage~\cite{Babichev:2009ee}. Simulations of non-vacuum spacetimes assuming spherical symmetry have been performed in Refs.~\cite{terHaar:2020xxb,Bezares:2021yek} and nonspherical oscillation of neutron stars in~\cite{Shibata:2022gec}, showing that the screening mechanism works in the case of stellar oscillation and suppresses the monopolar scalar emission. Furthermore, in collapsing stars scenarios~\cite{Bezares:2021yek}, the Cauchy problem, although locally well-posed, can lead to diverging characteristic speeds for the scalar field, which can be solved by introducing a fixing-the-equation technique~\cite{Cayuso:2017iqc}. A similar study about the chameleon screening mechanism has been done in Ref.~\cite{Dima:2021pwx}.

Alternative theories such as action~\eqref{actionst}, scalar-Gauss-Bonnet gravity, Ricci-Gauss-Bonnet gravity, and k-essence are all subclasses of the Horndeski theories, which are defined as the most general scalar-tensor theory that leads to second-order field equations. For a comprehensive and specialized review of Numerical Relativity and Horndeski theories, we refer the interested reader to Ref.~\cite{Ripley:2022cdh}.

\section{Exotic compact objects in binaries}\label{Section3}

Once the stability of isolated configurations has been addressed, the next step is to consider a more interesting scenario from the point of view of gravitational wave astronomy. It is widely known that most astrophysical systems involving compact objects are actually binaries, for which the orbit will shrink due to the emission of gravitational radiation, leading to the eventual collision. As for black holes and neutron stars, the only way to obtain precise gravitational waveforms emitted by binaries of exotic compact objects is to perform fully non-linear numerical relativity simulations of mergers of such objects. The available numerical techniques and code infrastructures that enable the evolution of black hole and neutron star binaries can also be used to some extent for some exotic objects, facilitating the task at hand. Indeed, in many cases, binary mergers have been successfully performed, albeit under some approximations, and their gravitational wave signatures are not completely unknown. However, even though academic studies that consider only a few specific configurations are very relevant and may point out the key differences with respect to black holes and neutron stars, they might not be sufficient for any potential discovery. To reliably confirm a detection of these objects would require carrying out hundreds of simulations, systematically varying the parameters that define the models under study, as it has been done for black holes and neutron stars, with the aim of building catalogs of gravitational waves that can be then compared to the observational data of the LIGO-Virgo-KAGRA detectors. 

\begin{svgraybox}Understanding and modelling the dynamics, endpoint, and gravitational wave emission of the different solutions is, therefore, necessary for the falsifiability of these alternative models. In this section, we will review some interesting cases of exotic binary simulations.\end{svgraybox}

\subsection{Bosonic stars in binaries}

There are no clear mechanisms as to how bosonic stars would condense and form during the early Universe. Anisotropies in the bosonic field could lead to incomplete gravitational collapse and star formation, through gravitational cooling~\cite{Seidel:1993zk}. Therefore, as for main sequence stars, it is likely that bosonic stars could come in binaries. Binary systems would eventually merge and emit a gravitational-wave signal in frequency and amplitude ranges that  directly depend on the masses of the stars and therefore on the mass of the bosonic particle $\mu$. 

\subsubsection{Initial data}

The initial method consists in taking a superposition of the two stars as initial data~\cite{pale1,pale2,2010PhDT.......390M,Bezares:2017mzk,Sanchis-Gual:2018oui}:
\begin{enumerate}
    \item Construct the initial data for two non-spinning boson stars with $\lbrace g^{(i)}_{ab},\Phi^{(i)}\rbrace$ and extend it to a Cartesian grid.
    \item Perform a Lorentz transformation in a chosen direction.
    \item Superpose the boosted solutions as:\begin{eqnarray}
        \Phi(t,r)&=&(\Phi^{(1)}_{0}(r)e^{-i(\omega t+\epsilon^{(1)})})_{\rm{boost}}+(\Phi^{(2)}_{0}(r)e^{-i(\omega t+\epsilon^{(2)})})_{\rm{boost}},\\
        g_{ab}&=&g_{ab, \rm{boost}}^{(1)}+g_{ab, \rm{boost}}^{(2)}-\eta_{ab}~,
    \end{eqnarray}
where $\epsilon^{(i)}$ is the initial phase of the stars and $\eta_{ab}$ is the Minkowski metric.
\end{enumerate}
This approach is not a proper solution to Einstein's equations but only a convenient approximation. Obtaining suitable initial data requires numerically solving four elliptic equations, the Hamiltonian and momentum constraints, which is a complex process in most cases. Recent efforts have been made to solve the binary boson-star initial data. In Refs.~\cite{Helfer:2018vtq,Helfer:2021brt,Evstafyeva:2022bpr}, the authors have improved the usual superposition of the single-star spacetime for the case of head-on collisions of equal and unequal mass, non-spinning boson star considering the potential for mini-boson stars and solitonic boson stars\footnote{Solitonic or non-topological boson stars are characterised by an hexic potential~\cite{1992PhR...221..251L,frie}.}. However, the constraint violation for the Hamiltonian and momentum constraint still needs to be eliminated. 

Newly, two groups have solved the initial data for boson star binaries in the last few months~\cite{Siemonsen:2023age,Atteneder:2023pge}. The first has developed a method for constructing binary boson star initial data that satisfies the constraints. The authors used the conformal thin-sandwich formalism, which had previously used initial data from black holes and neutron stars~\cite{East:2012zn}. They could construct initial data for various configurations considering different potentials, mass-ratios, spin magnitudes, and spin orientations. The last group that got accurate initial data was Ref.~\cite{Atteneder:2023pge}. Here, the authors constructed initial data also using the conformal thin-sandwich formalism and superposed free data equations for the head-on collision of mini-boson stars. They have numerically solved conformal thin-sandwich equations using the code \rm{bamps} and the hyperbolic relaxation method presented in Ref.~\cite{Ruter:2017iph}. Last but not least, one of the first papers in which the constraints have been appropriately solved was in Ref.~\cite{Dietrich:2018bvi}, focusing on head-on collisions of boson stars with neutron stars.

\subsubsection{Boson-Boson and Proca-Proca mergers}

\begin{figure}[h]
\sidecaption
\includegraphics[width=1\linewidth]{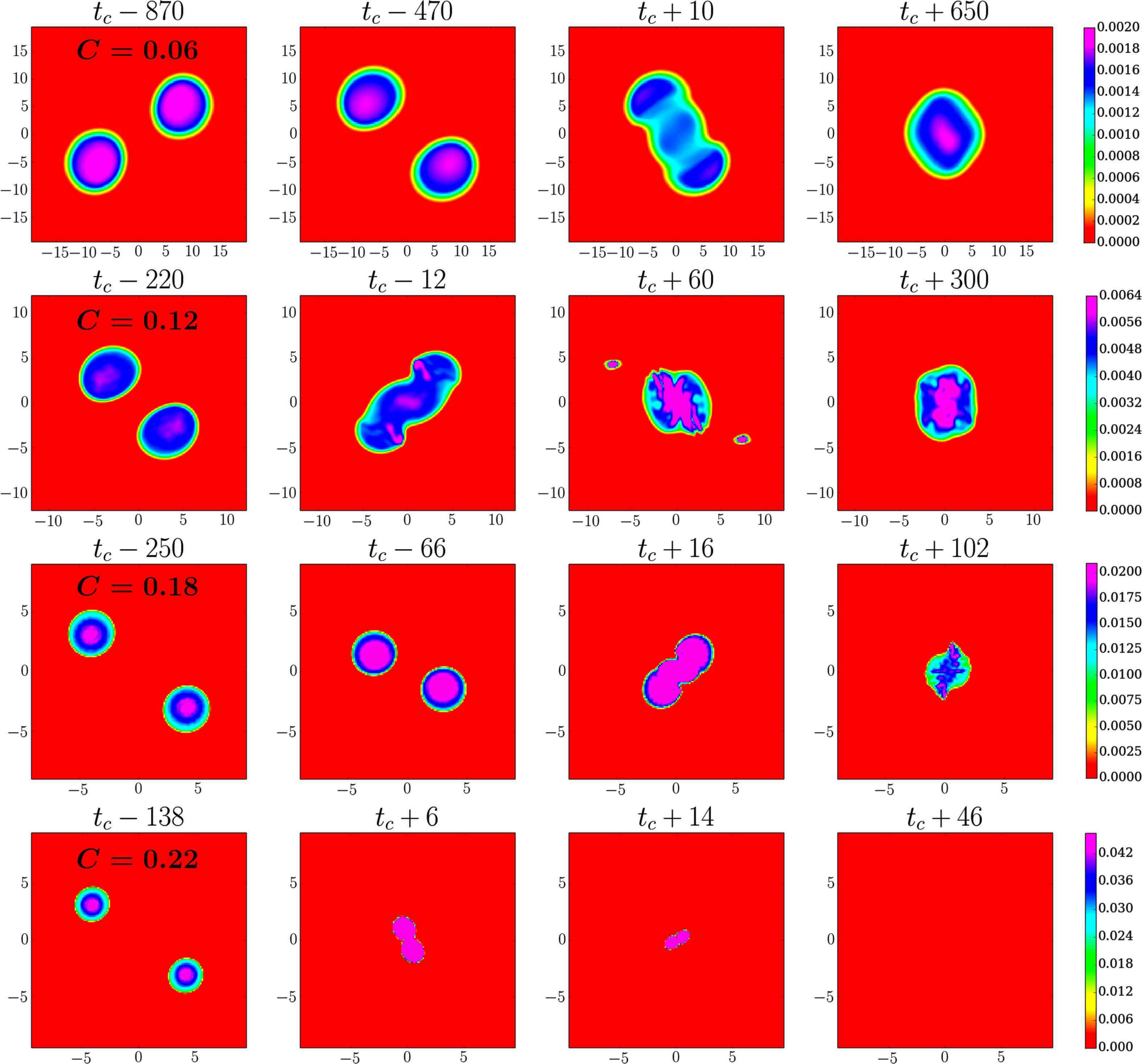}\hspace{0.01\linewidth}\\
\caption{ {\em Boson-Boson and Proca-Proca mergers.} Coalescence of binary boson stars: Snapshots in time of the Noether charge density in the orbital plane. Each row corresponds to a
different compactness (from top to bottom: 0.06, 0.12, 0.18, and 0.22). Extracted from~\cite{Bezares:2017mzk}.}
\label{fig:14211_bos}       
\end{figure}

In recent years, a significant number of numerical studies have been carried out on the collisions of bosonic stars and their emission of gravitational waves in General Relativity, with early simulations focusing on head-on collisions of non-spinning scalar boson stars.  We will begin this section by presenting a list of the main works related to this topic:

\begin{itemize}
    \item boson star head-on collisions~\cite{bernal2006scalar,pale1,Choptuik:2009ww,2010PhDT.......390M,Cardoso:2016oxy,Helfer:2021brt,Jaramillo:2022zwg,Atteneder:2023pge};
    \item boson star orbital mergers~\cite{pale2,Palenzuela:2017kcg,Bezares:2017mzk,Bezares:2018qwa,Sanchis-Gual:2020mzb,Bezares:2022obu,Croft:2022bxq,Evstafyeva:2022bpr,Siemonsen:2023age,Siemonsen:2023hko};
    \item oscillatons (real scalar fields)~\cite{Helfer:2018vtq,Widdicombe:2019woy};
    \item $\ell$-boson stars~\cite{Jaramillo:2022zwg};
    \item axionic boson stars~\cite{Sanchis-Gual:2022zsr};
    \item Proca stars~\cite{Sanchis-Gual:2018oui,Bustillo:2020syj,sanchis2022impact,CalderonBustillo:2022cja}.
\end{itemize}

Numerical simulations of bosonic star collisions have shown that their compactness $C\equiv M/R$ ($M$ and $R$ being the mass and radius of the boson star, the latter defined as the radius
within which $99\%$ of the total mass is contained) changes the outcome of the merger, the endpoint being a black hole or culminating in a more massive configuration~\cite{Palenzuela:2017kcg,Bezares:2017mzk,Sanchis-Gual:2018oui,Bezares:2022obu}, as illustrated in Fig.~\ref{fig:14211_bos}. 

Nevertheless, orbital mergers of non-spinning scalar boson stars unexpectedly failed to form rotating solutions of boson stars~\cite{Palenzuela:2017kcg,Bezares:2017mzk,Bezares:2022obu,Croft:2022bxq}, the final remnant losing all its angular momentum. This was later linked to the non-axisymmetric instabilities that appear when evolving isolated rotating boson stars~\cite{Sanchis-Gual:2019ljs, DiGiovanni:2020ror,Siemonsen:2020hcg,Dmitriev:2021utv}, making them quickly unstable by decaying to static spherically-symmetric configurations or collapsing to black holes. In the case of Proca stars, no rotating Proca star remnants were formed in the orbital collisions either~\cite{Sanchis-Gual:2018oui}, even though such rotating solutions are dynamically stable. It was recently found that the inability to retain the angular momentum is however related to the wave-like nature of bosonic fields and the quantized angular momentum given by the azimuthal winding number $m$ (see equation~\ref{eq:ansatz}). In particular, the interaction between the stars and the amount of energy released in the collision depend on the relative phases $\epsilon^{(i)}$ at merger between the objects~\cite{pale1,sanchis2022impact,Siemonsen:2023hko}. For instance, a phase difference of $\Delta\epsilon=|\epsilon^{(1)}-\epsilon^{(2)}|=\pi$ can lead to repulsive force between the stars that will keep them apart for a longer time. Refs.~\cite{pale1,Bezares:2017mzk} demonstrated that two non-rotating $\ell=m=0$ scalar boson stars with such phase difference of $\pi$ will bounce back and forth in a head-on collision and could eventually settle into a $\ell=1$, $ m=0$ dipole configuration~\cite{Ildefonso:2023qty} if self-interactions are considered (see also Ref.~\cite{Jaramillo:2022zwg} for the case of $\ell$-boson stars). Therefore, in mergers with orbital angular momentum, a rotating $\ell=m=1$ boson star can only be formed if the phase difference of the stars at merger is $\Delta\epsilon\gtrsim 7\pi/8$ as found in~\cite{Siemonsen:2023hko}. Self-interaction must also be included in order to stabilize the rotating scalar solutions.

\begin{figure}[h]
\sidecaption
\includegraphics[width=0.9\linewidth]{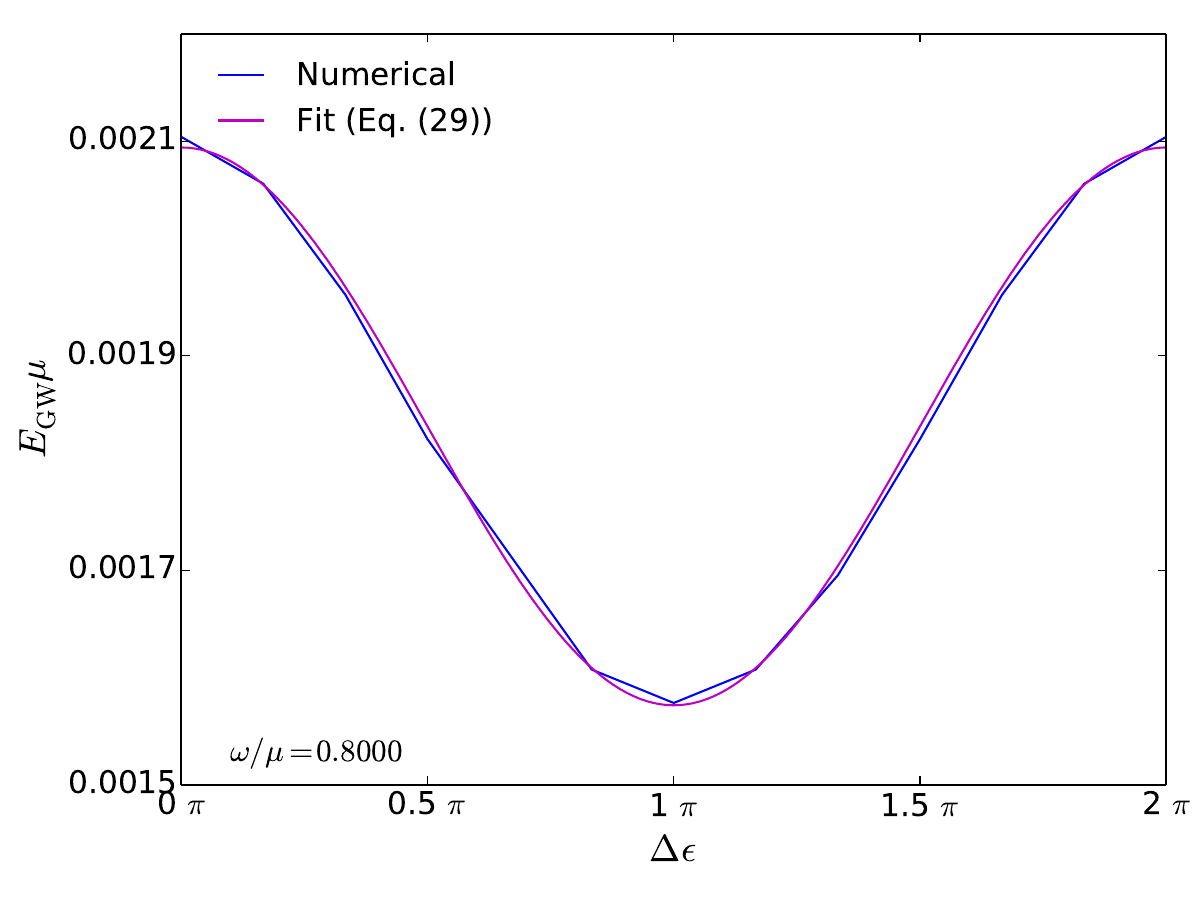}
\caption{ {\em Boson-Boson and Proca-Proca mergers.} Coalescence of binary boson stars: Gravitational-wave energy emitted in a head-on collision of equal-mass rotating Proca stars with fixed $\omega^{(1)}/\mu=\omega^{(2)}/\mu=0.8000$ and non-zero phase difference $\Delta\epsilon$ between the stars. The magenta line depicts the behaviour of the square of the Proca field amplitude as a function of $\Delta\epsilon$ computed from Eq.~(\ref{phase}). Extracted from~\cite{sanchis2022impact}.}
\label{fig:14212_energy}       
\end{figure}

The phase difference $\Delta\epsilon$ plays a key role in the outcome of the final configuration in the formation scenario when the stars are not compact enough to collapse. But it also has a crucial impact on the emission of gravitational waves. The wave-like nature of bosonic stars can lead to an interference at the time of the collision, greatly changing the amount of gravitational radiation and the morphology of the waveform~\cite{sanchis2022impact,Siemonsen:2023hko}. Varying the initial phases $\epsilon^{(i)}$ does not modify the initial energy density of the stars since it is independent of the phase. However, the value of $\Delta\epsilon$ at merger may cause a constructive or destructive interference pattern. From the amplitude of the Proca field $\mathcal{A}$, assuming a linear superposition of both stars $\mathcal{A}=\mathcal{A}^{(1)}+\mathcal{A}^{(2)}$ oscillating at with frequencies $\omega^{(1)}$ and $\omega^{(2)}$~\cite{sanchis2022impact} (see~\cite{pale1} for the scalar case):
\begin{eqnarray}\label{interference}
\rm{Re}(\mathcal{A})&\sim&\cos(\omega^{(1)}\, t + \epsilon^{(1)}) + \cos(\omega^{(2)}\, t+ \epsilon^{(2)})\\
&\sim&2\cos\biggl(\frac{(\omega^{(1)}+\omega^{(2)})}{2}t+\epsilon^{(1)}+\epsilon^{(2)}\biggl)\cos\biggl(\frac{(\omega^{(1)}-\omega^{(2)})}{2}t+\Delta\epsilon\biggl)\,,\nonumber\\
\rm{Im}(\mathcal{A})&\sim&\sin(\omega^{(1)}\, t+\epsilon^{(1)}) + \sin(\omega^{(2)}\, t+\epsilon^{(2)})\\
&\sim&2\sin\biggl(\frac{(\omega^{(1)}+\omega^{(2)})}{2}t+\epsilon^{(1)}+\epsilon^{(2)}\biggl)\cos\biggl(\frac{(\omega^{(1)}-\omega^{(2)})}{2}t+\Delta\epsilon\biggl)\nonumber.
\end{eqnarray}
The total magnitude of the field is then given by
\begin{eqnarray}\label{phase}
|\mathcal{A}|^2&=&{\rm Re}(\mathcal{A})^2+{\rm Im}(\mathcal{A})^2\sim4\cos^2\biggl(\frac{(\omega^{(1)}-\omega^{(2)})}{2}t + \Delta\epsilon\biggl)\nonumber\\
&\sim&2\bigl[1+\cos\bigl((\omega^{(1)}-\omega^{(2)})t+\Delta\epsilon\bigl)\bigl].
\end{eqnarray}
Eq.~(\ref{phase}) clearly indicates that at the time of the merger, the frequency difference $\Delta\omega$ and $\Delta\epsilon$ will change the interaction of Proca field describing both stars and induce a different gravitational wave emission depending on the value of $\cos\bigl((\omega^{(1)}-\omega^{(2)})t+\Delta\epsilon\bigl)$, in particular if $(\omega^{(1)}-\omega^{(2)})t+\Delta\epsilon=\pi$, the amplitude reaches its minimum value. Keeping fixed $\Delta\omega$, for instance in the case of equal-mass stars (for which $\omega^{(1)}=\omega^{(2)}$), the total energy emitted in gravitational waves $E_{\rm{GW}}$ in a head-on collision of rotating Proca stars greatly depends on $\Delta\epsilon$ as shown in Fig.~\ref{fig:14212_energy}. This result reveals the relevance of the phase difference and any study on merging bosonic stars must incorporate this parameter.

\begin{figure}[h]
\sidecaption
\includegraphics[width=0.8\textwidth]{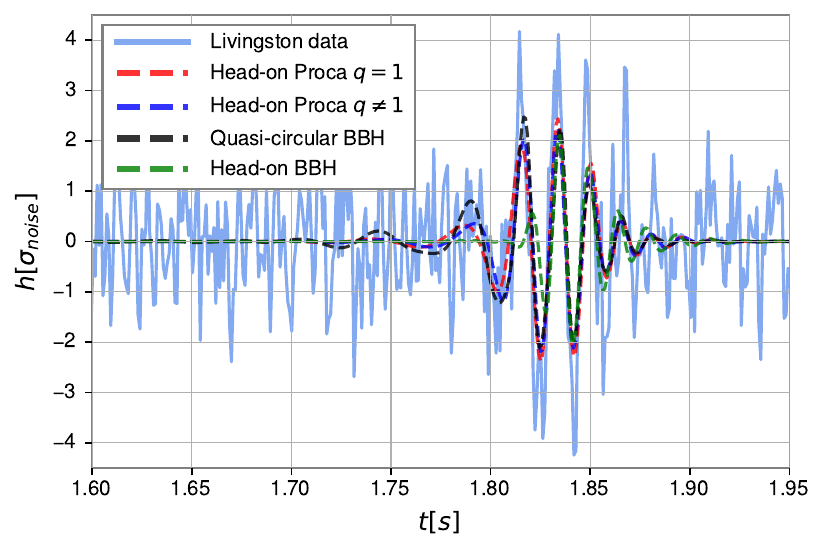}\\
\includegraphics[width=0.8\textwidth]{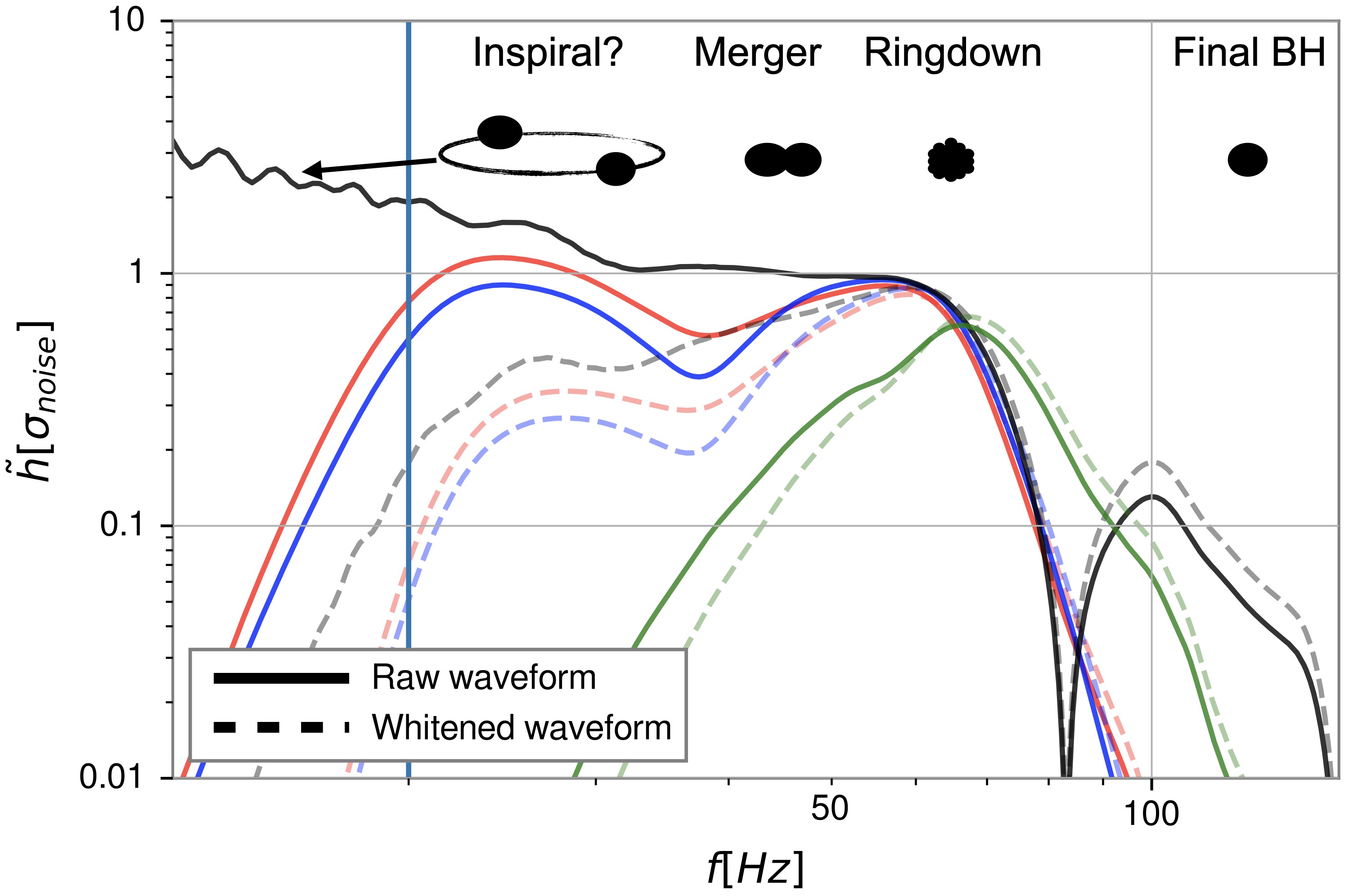}
\caption{ {\em Boson-Boson and Proca-Proca mergers.}  Time series and spectrum of GW190521. Top: Whitened strain data of the LIGO Livingston detector at the time of
GW190521, together with the best fitting waveforms for a head-on merger of two black holes (green), two equal or unequal mass Proca stars (red
and blue) and for a quasicircular black hole merger (black). Bottom: corresponding waveforms shown in the Fourier domain. Solid lines denote raw waveforms
(scaled by a suitable, common factor) while dashed lines show the whitened versions. The vertical line denotes the 20 Hz limit, below
which the detector noise increases dramatically. Extracted from~\cite{Bustillo:2020syj}.}
\label{fig:14213_gw}       
\end{figure}

Finally, highly compact bosonic star binaries need to be highlighted in the context of the current gravitational-wave detections by the LIGO-Virgo-KAGRA network. The compactness of those stars is such that it will lead to black hole collapse quickly after merger and to the formation of an event horizon, which means that a ringdown phase in the gravitational-wave signal will occur. In consequence, these systems can imitate part of the gravitational-wave phenomenology of black hole binary mergers~\cite{Bezares:2017mzk,Helfer:2018vtq,Sanchis-Gual:2018oui}, which are the astrophysical scenario that best fit so far the majority of the real gravitational-wave observations. In this regard, in~\cite{Bustillo:2020syj} a catalogue of around eighty numerical-relativity gravitational waveforms from head-on collision of rotating Proca stars (presented in~\cite{sanchis2022impact}) was compared using a Bayesian analysis with the gravitational wave event GW190521~\cite{LIGOScientific:2020iuh}. A slight statistical preference for the Proca model was found with respect to black hole binary mergers~\cite{Bustillo:2020syj} (see Fig.~\ref{fig:14213_gw}), although this does not constitute any strong evidence in favour of this alternative model, especially since head-on collisions are not expected to occur in the Universe. This particular event was selected due to it is larger total mass, which led to a final black hole with $150$ $M_\odot$, since it was observed as a short, burst-like signal which could be compatible with a head-on or eccentric collisions. Later, the same analysis was performed on this and a few other massive events with a larger set of waveforms from Proca stars including unequal-mass collisions in~\cite{sanchis2022impact,CalderonBustillo:2022cja}. While the preference was still there for a couple of the events, any further evidence remained inconclusive. An interesting point is that if the Proca hypothesis to explain similar events was found to be true, the mass of the hypothetical fundamental boson particle could be inferred through parameter estimation. As a proof of concept, in~\cite{Bustillo:2020syj,CalderonBustillo:2022cja} it was estimated that an ultralight vector boson of mass $\sim9\times10^{-13}$ eV could be the constituent of such Proca stars and potentially be part of the dark matter in the Universe.

\subsubsection{Fermion-boson -- fermion-boson mergers}
\begin{figure}[h!]
\centering
\includegraphics[width=0.5\linewidth]{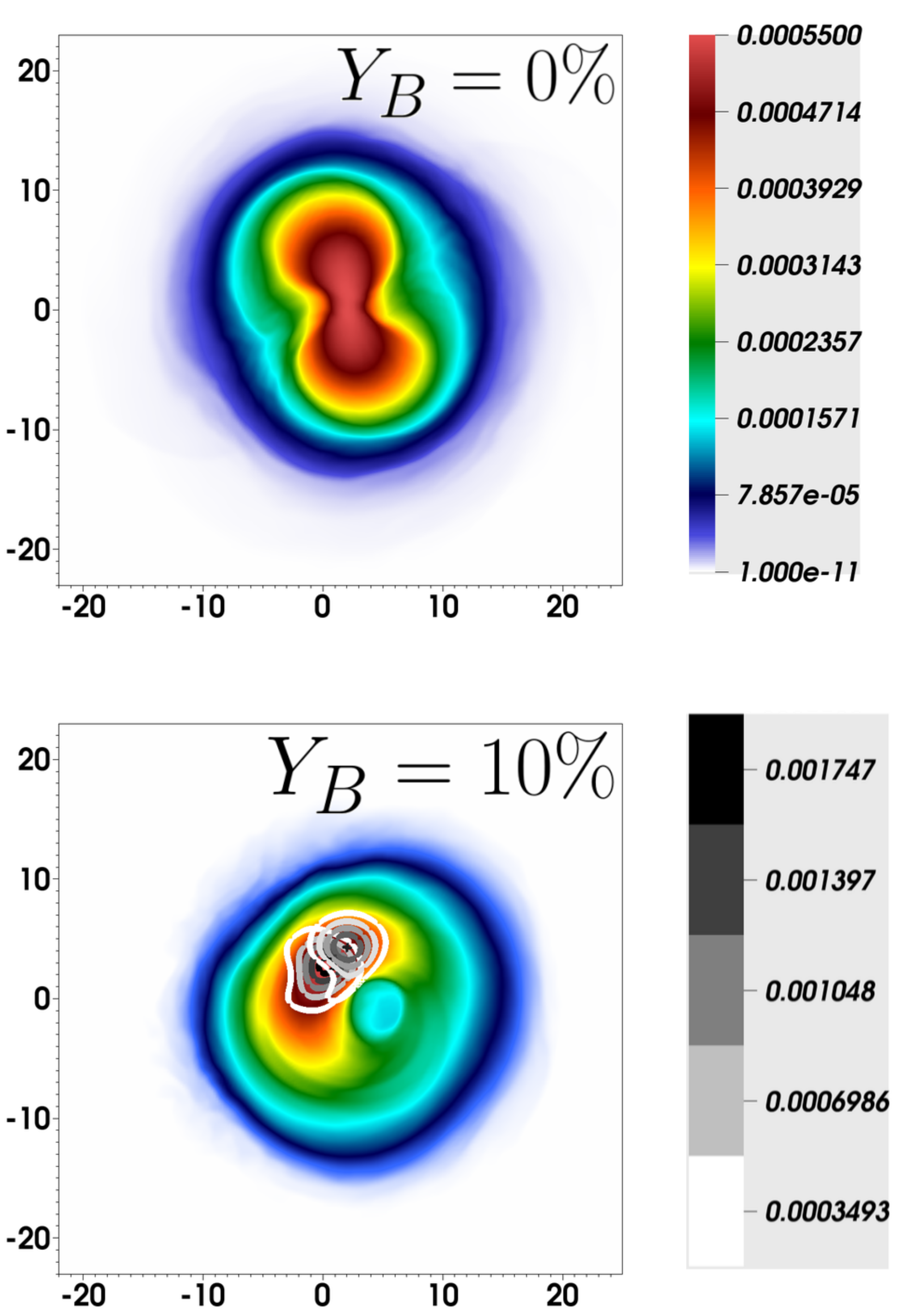}\hspace{-0.01\linewidth}
\includegraphics[width=0.48\linewidth]{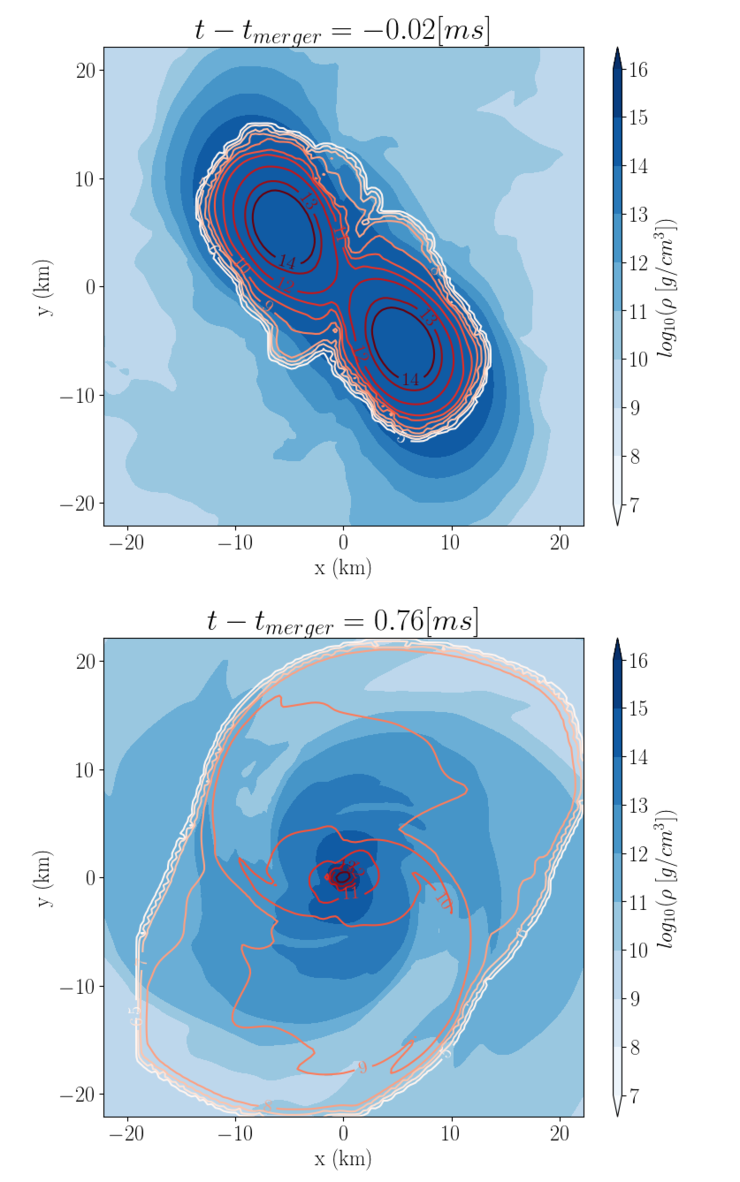}\hspace{-0.01\linewidth}
\caption{ {\em Fermion-boson -- fermion-boson mergers.} The left panels show the symmetry breaking of the quadrupole structure when there is a significant amount of bosonic component inside the neutron star in the post-merger stage. Extracted from~\cite{Bezares:2019jcb}.  The right panels show time snapshots in the equatorial plane of the different phases of the coalescence taken from the simulation of equal mass neutron stars and $5\%$ mirror dark matter. Extracted from~\cite{Emma:2022xjs}. }
\label{fbfb}
\end{figure}

As explained in the previous Section~\ref{fbs}, fermion-boson stars have been used as laboratories to study dark matter using neutron stars. There are models in which dark matter can form small clusters inside neutron stars~\cite{brito,Giangrandi:2022wht,Karkevandi:2021ygv}. These dark matter clusters can change the compactness and properties of stars; see Refs.~\cite{Baiotti:2019sew,Visinelli:2021uve} for a review. Therefore, collisions of these compact objects could play an important role in gravitational waves detection in binary neutron star merger events and shed some light on dark matter. Motivated by this, the first simulation of neutron star binaries with dark matter cores by numerical relativity simulations was done in Ref.~\cite{Bezares:2019jcb}. The authors have considered binaries of  fermion-boson stars, where the fermionic part models the neutron star and the bosonic part models the dark matter. In this case, equal-mass mergers of these stars with a fraction of dark matter in their interior form a one-arm instability during the post-merger (see Fig.~\ref{fbfb}), leaving an imprint on the emitted gravitational waves. In Ref.~\cite{Ellis:2017jgp}, a theoretical toy model of binary of neutron with dark matter in their interior has been considered, showing new frequencies in the post-merger signal. In the same context, the coalescence of binary neutron star admixed with dark matter has been studied in Ref.\cite{Bauswein:2020kor}, but the dark matter is considered a test particle. The dark matter particle has been modelled and simulated using smooth particle hydrodynamic methods. The authors have shown that dark matter affects the gravitational wave, especially in the post-merger. Here, different masses and equation-of-states for the neutron stars have been considered.

Furthermore, in Ref.~\cite{Emma:2022xjs}, the effects of mirror dark matter in neutron star mergers have been studied. The authors have extended the code BAM~\cite{Dietrich:2015iva} to perform this simulation, considering a two-fluid approach. They found that these binaries have a long inspiral and modified the ejecta and the disk mass, see Fig.~\ref{fbfb}. Finally,  consistent quasi-equilibrium configurations have been constructed for binary systems consisting of two neutron stars, each admixed with dark matter,  using the SGRID code in Ref.~\cite{Ruter:2023uzc}.

\subsubsection{Hybrid binaries mergers}

Recent numerical relativity simulations have also investigated the merger of hybrid binaries of boson stars and black hole/neutron stars. One of the first three-dimensional numerical relativity simulations of neutron star – boson star collisions was performed in Ref.~\cite{Dietrich:2018bvi}. These collisions can result in the formation of either a black hole or a neutron star surrounded by a bosonic cloud, with gravitational wave signals offering insights into the nature of the merger remnants. 

\begin{figure}[h]
\sidecaption
\includegraphics[width=1\linewidth]{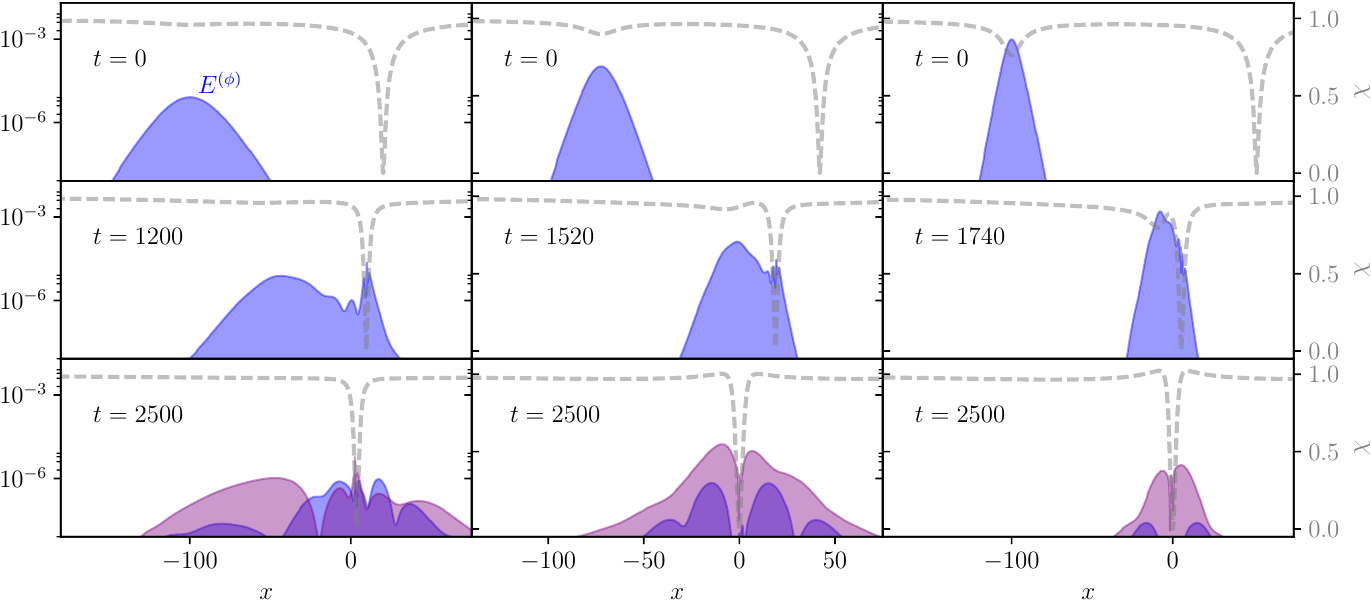}
\includegraphics[width=1\linewidth]{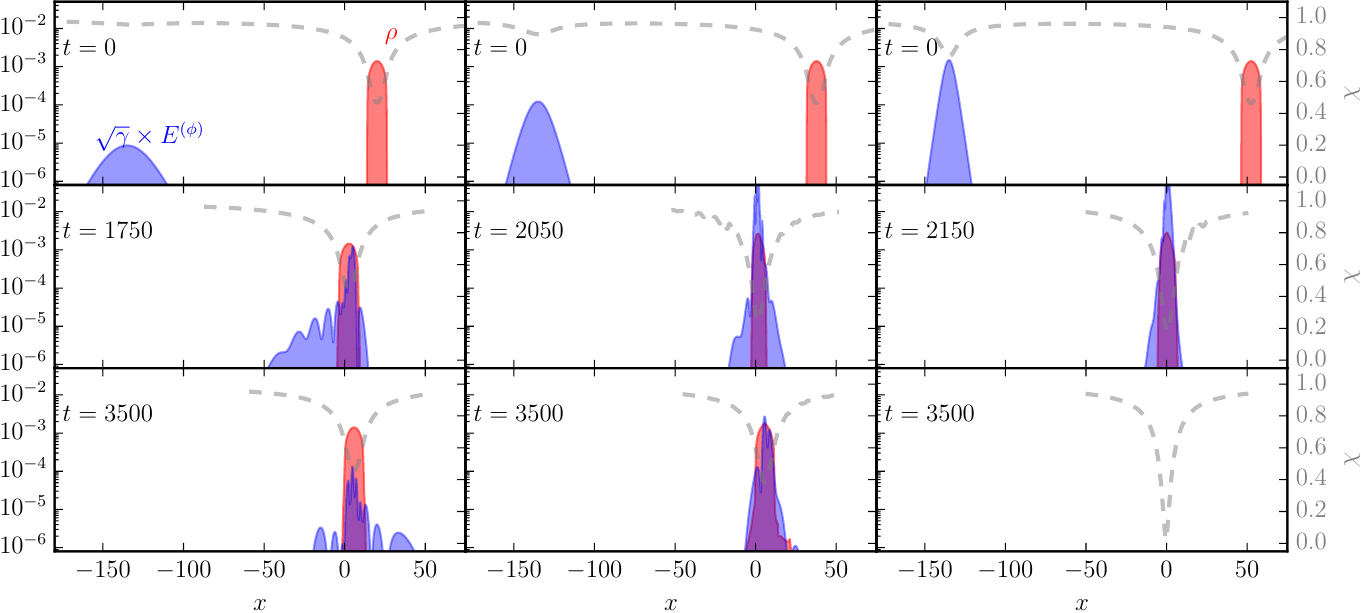}
\caption{ {\em Hybrid binaries mergers.} Time snapshots of the evolution of the bosonic energy density and the metric's conformal factor, which roughly represents the gravitational potential (grey) for different configurations of black hole—axion star collisions considered in Ref.~\cite{Clough:2018exo}. The bottom panel is the same as above but for neutron star—axion star collisions. The red colour represents the baryonic density. Extracted from~\cite{Clough:2018exo}.}
\label{bsbhns}       
\end{figure}

In Ref.~\cite{Clough:2018exo}, the authors have investigated the collision of axion stars with black holes and neutron stars, see Fig.~\ref{bsbhns}. These simulations have elucidated the potential role of axion star-black hole mergers in the formation of scalar hair around black holes. Similarly, in Refs.~\cite{Clough:2018exo, Dietrich:2018jov}, studies examining neutron star-axion have shown many observable phenomena across electromagnetic and gravitational wave channels. Moreover, in Ref.\cite{Dietrich:2019shr}, the authors explored if the role of nucleon-nucleon-axion bremsstrahlung has any insights into the cooling binary neutron star remnants. However, they found that the imprint of axions on gravitational wave signals and ejecta mass appears to be relatively small. 

Finally, for the case of the interaction between black holes and boson stars, simulations of the phenomenon of black holes piercing through massive bosonic stars have been investigated in Ref.~\cite{Cardoso:2022vpj}. The authors have challenging simulations involving black holes and boson stars, with length ratios as large as sixty-two. The authors have found that when the black hole pierces through the bosonic structure, the process is characterised by accretion-induced deceleration and the emission of gravitational waves. In Ref.~\cite{Cardoso:2022nzc}, in the context of fuzzy dark matter, the accretion of a boson star by a central black hole has been investigated in spherical symmetry. Finally, the authors in Ref.~\cite{Zhong:2023xll} have extended previous works by including a hexic solitonic potential $\Phi^{6}$ and exploring the piercing of a solitonic boson star by a boost black hole.

\subsection{Strange stars}\label{bnstrs}
As explained in Section~\ref{strange}, only a few studies have explored the dynamics of quark star binary systems. This is partially due to the numerical challenges that have to be faced when modelling the remarkable discontinuities on the surfaces of these stars. Here, we will only give a short overview of numerical relativity simulations of binary quark stars. 

In Ref.~\cite{Zhu:2021xlu}, the authors have performed quark-star binaries for the first time through fully general-relativistic simulations, displaying differences between them and their hadron star counterparts. The authors have found subtle differences in merger and post-merger frequencies in these simulations, particularly regarding tidal deformability and stellar compactness. Moreover, in Ref.~\cite{Zhou:2021tgo}, the post-merger of the evolution of equal-mass binary quark star coalescence has unveiled the possibility of prompt collapse to a black hole for supramassive merger remnants, challenging conventional post-merger scenarios observed in binary neutron star mergers. These findings emphasise quark-star binaries' distinctive dynamics and astrophysical implications, highlighting the need for further exploration in these simulations.

\subsection{Beyond General Relativity: black holes -black holes and   neutron star - neutron star mergers}\label{subsectionbGR}

To date, null and consistency tests have primarily validated the consistency of General Relativity in binary merger dynamics and in the strong-field regime. However, these tests are insufficient to rule out specific alternative theories of gravity. This is because accurate predictions for the merger waveforms of compact objects in these alternative theories have been obtained only in a limited number of cases. Such predictions necessitate the precise numerical solution of the theory's field equations, which involve complex systems of partial differential equations. A significant challenge in this area is the general difficulty in establishing well-posed initial data (Cauchy) problems for most alternative theories of gravity, as was explained in detail in Section~\ref{isolatedbgGR}.

Nevertheless, recent progress in numerical relativity simulations beyond General Relativity has been made~\cite{LISAConsortiumWaveformWorkingGroup:2023arg,Ripley:2022cdh}. Studies on binary black holes in scalar-tensor theories of gravity, such as those by~\cite{Healy:2011ef,Berti:2013gfa}, have been among the first. Building on this, \cite{Cao:2013osa} examined binary black hole mergers within the $f(R)$ theory framework, finding distinct merger dynamics compared to the General Relativity counterpart. Further investigation on dynamical Chern-Simons gravity by~\cite{Okounkova:2017yby,Okounkova:2019dfo,Okounkova:2019zjf} have provided insights into the gravitational wave signatures produced by these mergers using an order-by-order expansion scheme. Notably, this technique can lead to secular growing solutions, breaking the perturbation theory on long timescales. However, this could be solved by relying on the numerical renormalization-group-based technique presented in Ref.~\cite{GalvezGhersi:2021sxs}.

Additionally, progress has been made in understanding neutron star mergers within these alternative frameworks. In Refs.~\cite{Barausse:2012da,Palenzuela:2013hsa} scalar-tensor theories were considered as the underlying theory of gravity, uncovering phenomena such as \textit{dynamical scalarization}, i.e., once neutron stars come close enough to each other, they dynamically scalarize. These studies were extended in Ref.~\cite{Shibata:2013pra}. In Ref.~\cite{Bezares:2021dma}, kinetic screening in simulations of merging binary neutron stars was studied, finding that the screening tends to suppress the (subdominant) dipole scalar emission but not the (dominant) quadrupolar scalar mode. 

\begin{figure}[h]
\sidecaption
\includegraphics[width=1\linewidth]{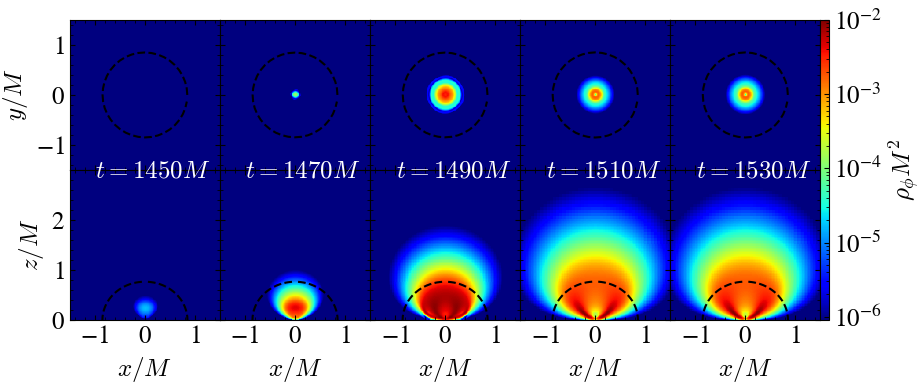}
\caption{{\em Beyond General Relativity: black holes - black holes and neutron star - neutron star mergers}. Time evolution of the scalar cloud after the merger of black holes binary in Einstein-scalar-Gauss-Bonnet gravity. Extracted from~\cite{AresteSalo:2023mmd}.}
\label{bGRbin}       
\end{figure}

The evolution of Einstein-scalar-Gauss-Bonnet gravity using modified harmonic formulations has also been a focus of attention~\cite{East:2020hgw,East:2021bqk,Corman:2022xqg}. In this context, Refs.~\cite{AresteSalo:2022hua,AresteSalo:2023mmd,Doneva:2023oww} have addressed the well-posedness of four-derivative scalar-tensor theories (including Einstein-scalar-Gauss-Bonnet gravity) by employing singularity avoiding coordinates and the CCZ4 formalism, see Fig~\ref{bGRbin}. Simulations in the decoupled limit aiming to study dynamical scalarization and descalarization of binary black holes have been conducted in Refs.~\cite{Benkel:2016kcq,Benkel:2016rlz,Silva:2020omi,Doneva:2022byd,Elley:2022ept}.  Finally, simulations of binary neutron star mergers in scalar-Gauss-Bonnet gravity have also been performed in Ref.~\cite{East:2022rqi}, discovering that, although there is almost no impact during the inspiral stage, in the post-merger phase; however, strong scalar effects appear. Last but not least, black hole binaries in cubic Horndeski theories with a massive potential have been investigated in Ref.\cite{Figueras:2021abd}.

\newpage


\begin{acknowledgement}
We thank S. Liebling and C. Herdeiro for their useful comments on the manuscript and N. Franchini for their enlightening conversations on the alternative theory of gravity and numerical relativity. NSG acknowledges support from the Spanish Ministry of Science and Innovation via the Ram\'on y Cajal programme (grant RYC2022-037424-I), funded by MCIN/AEI/10.13039/501100011033 and by ``ESF Investing in your future”. NSG is further supported by the Spanish Agencia Estatal de Investigaci\'on (Grant PID2021-125485NB-C21) funded by MCIN/AEI/10.13039/501100011033 and ERDF A way of making Europe, and by the European Horizon Europe staff exchange (SE) programme HORIZON-MSCA2021-SE-01 Grant No. NewFunFiCO-101086251.
\end{acknowledgement}






\bibliographystyle{myspringer}
\bibliography{biblio}

\end{document}